\documentclass[]{aa}

\usepackage{natbib}
\bibpunct{(}{)}{;}{a}{}{,} 
\usepackage{graphicx}
\usepackage{txfonts}

\usepackage{color}
\usepackage{threeparttable}
\usepackage{multirow}

\begin{document}

   \title{Recurring tidal disruption events a decade apart in IRAS F01004-2237}


   \author{Luming Sun\inst{1},
           Ning Jiang\inst{2,3},
          Liming Dou\inst{4},
          Xinwen Shu\inst{1},
          Jiazheng Zhu \inst{2,3},
          Subo Dong\inst{5,6,7},
          David Buckley\inst{8},
          S. Bradley Cenko\inst{9},
          Xiaohui Fan\inst{10},
          Mariusz Gromadzki\inst{11},
          Zhu Liu\inst{12},
          Jianguo Wang\inst{13},
          Tinggui Wang\inst{2,3},
          Yibo Wang\inst{2,3},
          Tao Wu\inst{1},
          Lei Yang\inst{1},
          Fabao Zhang\inst{1},
          Wenjie Zhang\inst{7},
          and Xiaer Zhang\inst{2,3}
          }
   \authorrunning{L.-M. Sun et al.}

   \institute{Department of Physics, Anhui Normal University, Wuhu, Anhui 241002, China. \\
             \email{sunluming@ahnu.edu.cn}
         \and
             CAS Key Laboratory for Researches in Galaxies and Cosmology, Department of Astronomy, University of Science and Technology of China, Hefei, Anhui 230026, China\\
             \email{jnac@ustc.edu.cn}
         \and
             School of Astronomy and Space Sciences, University of Science and Technology of China, Hefei, 230026, China
         \and
             Center for Astrophysics, Guangzhou University, Guangzhou 510006, China
         \and
             Department of Astronomy, School of Physics, Peking University, 5 Yiheyuan Road, Haidian District, Beijing 100871, China
         \and
             Kavli Institute of Astronomy and Astrophysics, Peking University, 5 Yiheyuan Road, Haidian District, Beijing 100871, China
         \and
             National Astronomical Observatories, Chinese Academy of Science, 20A Datun Road, Chaoyang District, Beijing 100101, China
         \and
             South African Astronomical Observatory, P.O. Box 9, Observatory 7935, Cape Town, South Africa
         \and
             Astrophysics Science Division, NASA Goddard Space Flight Center, MC 661, Greenbelt, MD 20771, USA
         \and
             Steward Observatory, University of Arizona, 933 North Cherry Avenue, Rm. N204, Tucson, AZ 85721-0065, USA
         \and
             Astronomical Observatory, University of Warsaw, Al. Ujazdowskie 4, 00-478 Warszawa, Poland
         \and
             Max-Planck-Institut f\"{u}r extraterrestrische Physik, Gie{\ss}enbachstra{\ss}e 1, 85748 Garching, Germany
         \and
             Key Laboratory for the Structure and Evolution of Celestial Objects, Yunnan Observatories, Kunming 650011, China
             }

   \date{Received XXX; accepted YYY}


  \abstract
   {In theory, recurring tidal disruption events (TDEs) may occur when a close stellar binary encounters a supermassive black hole, if one star is captured and undergoes repeating partial TDEs, or if both stars are tidally disrupted (double TDEs).
   In addition, independent TDEs may be observed over decades in some special galaxies where the TDE rate is extremely high.
   Exploring the diversity of recurring TDEs and probing their natures with rich observational data help to understand these mechanisms.}
   {We report the discovery of a second optical flare that occurred in September 2021 in IRAS F01004-2237, where the first flare occurred in 2010 has been reported, and present a detailed analysis of multi-band data.
   We aim to understand the nature of the flare and explore the possible causes of the recurring flares.}
   {We describe our analysis of the position of the flare, the multi-band light curves (LCs), the optical and ultraviolet (UV) spectra, and the X-ray LC and spectra.}
   {The position of the flare coincides with the galaxy centre with a precision of 650 pc.
   The flare peaks in $\sim50$ days with an absolute magnitude of $\sim-21$ and fades in two years roughly following $L\propto t^{-5/3}$.
   It maintains a nearly constant blackbody temperature of $\sim$22,000 K in the late time.
   Its optical and UV spectra show hydrogen and helium broad emission lines with full width at half maxima of 7,000--21,000 km s$^{-1}$ and He II/H$\alpha$ ratio of 0.3--2.3.
   It shows weak X-ray emission relative to UV emission, with X-ray flares lasting for $<2-3$ weeks, during which the spectrum is soft with a power-law index $\Gamma=4.4^{+1.4}_{-1.3}$.
   These characters are consistent with a TDE, ruling out the possibilities of a supernova or an active galactic nuclei flare.
   With a TDE model, we infer a peak UV luminosity of $3.3\pm0.2\times10^{44}$ erg s$^{-1}$ and an energy budget of $4.5\pm0.2\times10^{51}$ erg.}
   {A TDE caused the flare that occurred in 2021.
   The two optical flares separated by $10.3\pm0.3$ years can be interpreted as repeating partial TDEs, double TDEs, or two independent TDEs.
   Although no definitive conclusion can be drawn, the partial TDEs interpretation predicts a third flare around 2033, and the independent TDEs interpretation predicts a high TDE rate of $\gtrsim10^{-2}$ yr$^{-1}$ in F01004-2237, both of which can be tested by future observations.}


   \maketitle
%

\section{Introduction}

A tidal disruption event (TDE) occurs when a star wanders close enough to a supermassive black hole (SMBH) that the pericenter distance $r_p$ less than the tidal radius \citep[e.g.,][]{Rees1988}:
\begin{equation}
r_t \approx R_\star \left( \frac{M_{\rm BH}}{M_\star} \right)^{1/3},
\end{equation}
where $R_\star$ and $M_\star$ are the radius and the mass of the star, and $M_{\rm BH}$ is the mass of the SMBH.
TDEs produce intense electromagnetic radiation converted from the gravitational energy released by the accreted stellar debris, causing bright flares in X-ray, UV, and optical bands peaking in 1--2 months and fading in months to years in centers of galaxies \citep[e.g.,][]{Bade1996,Gezari2006,vanVelzen2011}.
In recent years, researchers have begun to study the diversity of the TDE process by considering partial TDEs and double TDEs.
Numerical simulations \citep[e.g.,][]{Guillochon2013,Ryu2020c} indicated that the outcome of a star encountering an SMBH depends on the parameter $\beta \equiv r_t/r_p$: the star is entirely disrupted when $\beta$ is above a critical value, depending on the stellar structure, and loses a fraction of mass for lower $\beta$, in which the fraction decreases continuously with decreasing $\beta$.
Theories predicted that partial TDEs are more common than full TDEs \citep{Stone2016a,Zhong2022}, and the LCs of partial TDEs decline more rapidly than that of full TDEs in the late-time \citep{Coughlin2019,Miles2020,Chen2021}.
If the stellar remnant is in an elliptical orbit after a partial TDE, it may be disrupted again after returning to the pericentre, resulting in repeating partial TDEs that are observed as periodic flares.
Periodic partial TDEs can result from the encounter of a close binary and an SMBH \citep{Cufari2022,LiuC2023}.
Under the Hills mechanism, in which one star in the binary is captured by the SMBH and the other ejected \citep[e.g.,][]{Hills1988}, if the captured star's pericenter is close enough to the SMBH, or if it is bounced by other existing captured stars into an orbit that facilitate disruption \citep{Bromley2012}, periodic partial TDEs can occur.
In addition, simulations show that the encounter of a close binary and an SMBH may lead to various outcomes.
One fascinating outcome is double TDEs, where both stars are tidally disrupted \citep{Mandel2015}.
Taking partial disruption into account, the possible outcomes may be even more diverse, and a full TDE and repeating partial TDEs may both occur \citep{Mainetti2016}.

Observationally, verification of a TDE is sometimes challenging because it needs to be distinguished from other TDE analogues \citep{Zabludoff2021}, such as superluminous supernovae \citep[SLSNe,][]{Gal-Yam2019} and active galactic nuclei (AGN) flares \citep[e.g.,][]{Kankare2017,Oknyansky2019,Frederick2019}.
Characteristics that distinguish TDEs from these analogues include positions consistent with the galactic nuclei, smooth power-law declines in the light curve (LC) roughly following $t^{-5/3}$, blackbody continuum in UV/optical bands with constant temperatures of $1.2-4\times10^4$ K for hundreds of days, broad emission lines (BELs) with widths remaining relatively broad over months to years, strong He II $\lambda$4686 BELs and sometimes nitrogen Bowen fluorescence (BF) lines, and soft X-ray emissions with steep spectra \citep[e.g.,][]{Saxton2020,vanVelzen2020,Charalampopoulos2022}.
It is generally necessary to assemble a variety of features to classify a flare.
However, even so, some flares, for example, CSS100217:102913+404220 \citep{Drake2011} and ASASSN-18jd \citep{Neustadt2020}, cannot be classified without doubt and are still referred to as ambiguous nuclear transients.

Till now, dozens of spectroscopically confirmed TDEs have been discovered, thanks to large-area high-cadence optical surveys such as the All-Sky Automated Survey for Supernovae \citep[ASASSN,][]{Shappee2014,Holoien2016_14li}, the Asteroid Terrestrial-impact Last Alert System \citep[ATLAS,][]{Tonry2018,Smith2020}, and the Zwicky Transient Facility \citep[ZTF,][]{vanVelzen2021,Hammerstein2023}.
With the large number, an accurate rate of optically selected TDE can be measured to be $3.1^{+0.6}_{-1.0}\times10^{-7}$ Mpc$^{-3}$ yr$^{-1}$, or $3.2^{+0.8}_{-0.6}\times10^{-5}$ galaxy$^{-1}$ yr$^{-1}$ for galaxies with $M_{\rm gal}\sim10^{10}$ $M_\odot$ \citep{Yao2023}.
The TDE rate varies greatly among different types of galaxies.
For example, the rate in post-starburst galaxies is higher than that in normal galaxies by a factor of several to several tens \citep{French2016,Graur2018,Yao2023}.
There are many possible explanations for this \citep[see][for a review]{French2020,Stone2020}, including SMBH binaries \citep[e.g.,][]{Ivanov2005}, central overdensities \citep[e.g.,][]{Stone2016a,Stone2016b}, and others.
These explanations facilitate a natural hypothesis that starburst galaxies would also have a high TDE rate.
\cite{Tadhunter2017} claimed to verify this hypothesis by finding a TDE candidate in 2010 in IRAS F01004-2237 (hereafter F01004-2237), during monitoring of a sample of 15 dusty starburst galaxies over a decade.
However, the result of \cite{Tadhunter2017} is controversial, as \cite{Trakhtenbrot2019} considered the TDE candidate in F01004-2237 as an AGN flare, and we will describe their debates in detail later.
The hypothesis of high TDE rate in starburst galaxies has no evidence from other optical surveys, possibly because these TDEs are obscured by dust \citep{Roth2021}.
This was supported by the discovery of obscured TDEs via mid-infrared flares such as Arp 299-B AT1 \citep[e.g.,][]{Mattila2018,Reynolds2022}.

Recently, optical rebrightenings or recurring flares have been found in a significant fraction of TDEs \citep[e.g.,][]{Jiang2019,Yao2023}.
Some of the recurring flares are suggested to be possibly associated with repeating partial TDEs.
Examples are: 1. the two X-ray flares observed in 1990 and 2010 in the Seyfert galaxy IC 3599 \citep{Grupe1995,Grupe2015,Campana2015};
2. the 114-day periodic flares in the galaxy ESO 253-G003 repeating for $>$21 times since the first detection in 2014 \citep{Payne2021,Payne2022,Payne2023};
3. the X-ray transient eRASSt J045650.3-203750 with five gradually weakening flares with time intervals of $\sim$200 days \citep{Liu2023,Liu2024};
4. the optical TDE AT 2018fyk and the faint optical flare 3.3 yr after it \citep{Wevers2019,Wevers2023};
5. the X-ray TDE candidate RX J133157.6-324319.7 in 1993 and the second X-ray flare in 2022 \citep{Hampel2022,Malyali2023};
6. the optical TDE AT 2020vdq and the second flare 2.7 yr after it \citep{Hammerstein2023,Somalwar2023}.
7. the optical TDE AT 2022dbl and the second flare 2 yr after it \citep{Lin2024}.
8. the optical flare AT 2019aalc in a Seyfert galaxy and the second flare 4 yr after it \citep{vanVelzen2024,MilanVeres2024}.
Here we do not include events with a double-peak light curve like AT 2019avd \citep{Malyali2021}, because the two peaks can be interpreted as different phases of the same TDE \citep{Chen2022}.
The characters of the recurring flares, including periods, luminosities, energy budgets, and SEDs, present colossal diversity.
Whether the diversity can be uniformly interpreted with repeating partial TDEs is still unclear.
In six out of the eight cases listed above, periodicity cannot yet be confirmed as only two flares were detected.
Moreover, there are alternative interpretations; for example, the X-ray flares in IC 3599 could be caused by extreme AGN disk instability \citep{Saxton2015}.
In addition, repeating flares with much lower energy budgets were observed in the X-ray band, known as quasi-periodic eruptions \citep[QPE,][]{Miniutti2019,Giustini2020,Arcodia2021,Evans2023}.
These QPEs occur with intervals of 2 hours to 1 month, releasing energies in the X-ray band of $\sim10^{45}-10^{48}$ erg with peak luminosity of $\sim10^{42}-10^{43}$ erg s$^{-1}$.
It is unclear how they are connected to the energetic recurring flares.
Further exploring the diversity of recurring flares and probing their natures with rich observational data help understand repeating partial TDEs and other mechanisms contributing to recurring flares.

In this work, we report on the optical flare in F01004-2237 that recurred in 2021, adding a new member to recurring flares.
\cite{Tadhunter2017} identified the first flare occurring in 2010 with V-band LC from Catalina Real-Time Transient Survey \citep[CRTS,][]{Drake2009}.
They demonstrated that the flare was caused by a TDE rather than a supernova and an AGN flare, using the WHT spectrum taken in September 2015, $\sim$5 years after the flare began.
The spectrum shows a broad He II $\lambda$4686 emission line with FWHM$\sim$6200 km s$^{-1}$, which was not seen in spectra taken before the flare.
Meanwhile, the new H$\beta$ emission is relatively weak with He II/H$\beta=5.3\pm0.9$.
These features are inconsistent with SNe and AGN flares and support a TDE interpretation.
However, \cite{Trakhtenbrot2019} questioned the TDE classification on the ground that they considered the spectrum of F01004-2237 to be similar to Bowen fluorescence flares in AGNs.
After the flare faded, we noticed another flare that occurred on September 4, 2021, from ATLAS and ASASSN data.
Hereafter, we will refer to the two flares as the 2010 and 2021 flares, respectively, based on the time of discovery.
The 2021 flare was also detected by the \textit{Gaia} space telescope \citep{Hodgkin2021} on January 21, 2022, with the name Gaia22ahd ($=$AT2022agi).
In this paper, we describe the observations and data reductions in section 2, present our analysis of the multi-band data in section 3, demonstrate the nature of the 2021 flare in section 4, explore the possible origins of the recurring flares in section 5, and discuss and summarise our results in section 6 and 7.
Thoughout this paper, we adopted $z=0.11783$ for F01004-2237 following \cite{Spence2018}, and a $\Lambda$CDM cosmology with $H_0=70$ km s$^{-1}$ Mpc$^{-1}$, $\Omega_M=0.3$ and $\Omega_\Lambda=0.7$, resulting in a luminosity distance of 548.8 Mpc.
All the UV/optical data were corrected for galactic extinction using $A_V=0.048$ from \cite{SFD2011}.

\section{Observations and data}

\subsection{Optical light curve data} \label{sec:2.1}

The ATLAS has been observing F01004-2237 in the $c$ and $o$ bands since August 2015.
We obtained the photometries on the subtracted images from the forced photometry server\footnote{https://fallingstar-data.com/forcedphot/} \citep{Shingles2021}.
We only selected data taken under good weather conditions with the criterion that the limit magnitude is fainter than 18 for both bands.
We then binned the LC data nightly\footnote{The survey strategy of ATLAS results in 2--8 consecutive exposures per night of observation.}.
The flux in the binned LC was calculated using the weighted average value of the single-exposure fluxes in the time bin, where the weight is the reciprocal of the square of the flux errors, and its error was calculated according to the law of propagation of uncertainties.
We finally obtained reference fluxes from the forced photometry server, 249 and 329 $\mu$Jy for $c$ and $o$ bands, respectively, and added them to the photometries.

The ASASSN has been observing F01004-2237 in the $g$ band since September 2017\footnote{Although ASASSN observed F01004-2237 in the $V$ band between 2013 and 2018, we did not use the data because the V-band observation covered neither of the two flares and the data quality is poor.}.
We obtained the original light curve data from the ASASSN Sky Patrol\footnote{https://asas-sn.osu.edu/} \citep{Kochanek2017}.
We used the photometries on the subtracted images with reference flux added and selected good data with a criterion that the limit magnitude is fainter than 18.
We binned the LC data nightly as we did for the ATLAS data.

We collected the $G$-band LC of F01004-2237 from the Gaia Photometric Science Alerts website\footnote{http://gsaweb.ast.cam.ac.uk/alerts/alert/Gaia22ahd/}.
We did not analyze this LC data in detail and only used it for display purposes because of the sparse sampling and the lack of error data.
To study the 2010 flare, we also collected the CRTS $V$-band LC of F01004-2237 from July 2005 to January 2014 from the data release 2\footnote{http://nesssi.cacr.caltech.edu/DataRelease/}.
CRTS observed the source 4--10 nights per year by taking 2--4 exposures per night, and we also binned the LC data nightly.

\begin{figure*}
\centering
  \includegraphics[width=17cm]{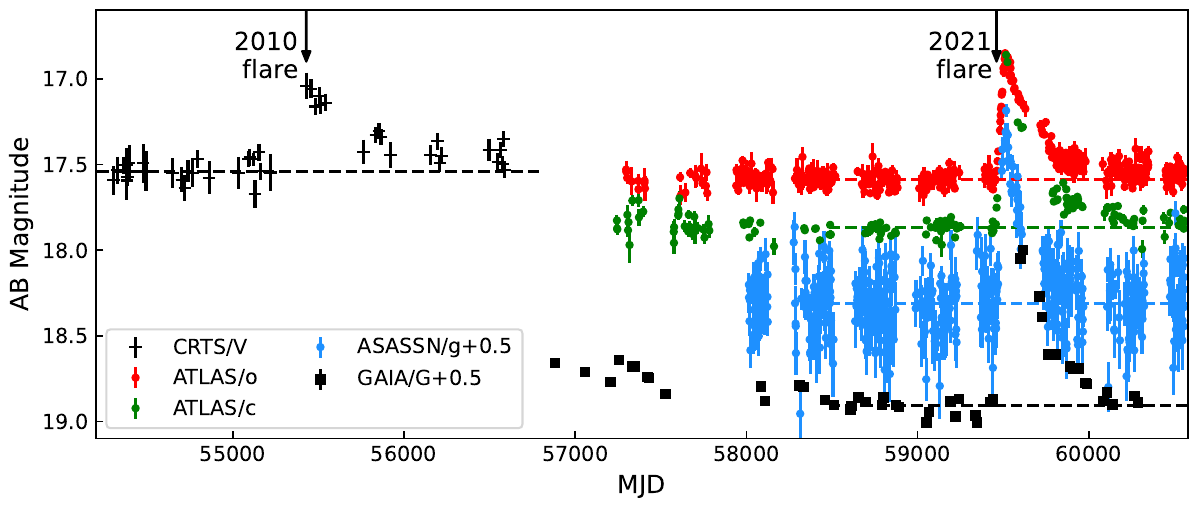}
  \caption{
  Long-term variability of F01004-2237 showing the two flares which occurred in 2010 and 2021, respectively.
  We show the levels before the flare occurred with horizontal dashed lines in corresponding colours and label the dates when the two flares were first detected.
  }
\label{fig:lc_opt}
\end{figure*}

We made follow-up observations in $g$, $r$ and $i$ bands from December 2021 to January 2023 using the Las Cumbres Observatory Global Telescope Network \citep[LCOGT,][]{Brown2013}.
Using the images taken before the flare by the SkyMapper Southern Survey \citep{skymapperDR1} as the reference images, we made image subtraction with the software Saccadic Fast Fourier Transform \citep[SFFT,][]{SFFTpaper}.
We then performed PSF photometries on the difference images using the \textit{Photutils} package of Astropy.
We finally made flux calibration using nearby stars with the source catalog\footnote{https://catalogs.mast.stsci.edu/panstarrs/ } of the Panoramic Survey Telescope \& Rapid Response System \citep[PanSTARRS,][]{PS1paper}.
We show all the original light curve data in Fig.~\ref{fig:lc_opt}.

\subsection{Follow-up \textit{Swift} and \textit{XMM-Newton} observations} \label{sec:2.2}

After we noticed the 2021 flare, we triggered target of opportunity observations of \textit{Swift} telescope.
We present the information of all the observations in Table~\ref{tab:xrtdata} in the appendix.

We obtained the XRT count rate in 0.3--10 keV of F01004-2237 from the UK Swift Science Data Centre\footnote{https://www.swift.ac.uk/user\_objects/} (UKSSDC).
Their server calculated the count rate following \citep{Evans2007,Evans2009} as:
\begin{equation}
r = \frac{(N_{\rm tot}-N_{\rm bkg})}{t_{\rm exp}}\times f_c
\end{equation}
where $N_{\rm tot}$ is the total photon count in the source region, $N_{\rm bkg}$ is the expected count of background photons in the source region, $t_{\rm exp}$ is the exposure time, and $f_c$ is the correction factor for pile-up, dead zones on the CCD, and source photons falling outside of the source extraction region.
We list the count rates with 1$\sigma$ errors (or 3$\sigma$ upper limits) and data for calculating them in Table~\ref{tab:xrtdata}.

We also extracted the photometry of F01004-2237 on the UVOT images using \textit{uvotsource} in High Energy Astrophysics Software (HEASoft).
We used an aperture with a radius of 5$\arcsec$, and calculated background in an annular region with inner and outer radii of 35$\arcsec$ and 40$\arcsec$, respectively.
The results are listed in Table~\ref{tab:uvotdata}.

We triggered a target of opportunity observation of \textit{XMM-Newton} telescope (ID 0911791101) on June 8, 2022.
In addition, there is an archival \textit{XMM-Newton} observation (ID 0605880101) on December 8, 2009 \citep[see details in][]{Nardini2011}.
This observation was taken before the 2010 flare occurred (later than Jan 2010; see section 3.2) and can be used to measure pre-flare fluxes from the host galaxy.
So we analyze both the two data.
We process the \textit{XMM-Newton} data using the Science Analysis System (SAS) version 20.0.0.
We included event patterns 0--4 for the PN camara and 0--12 for the MOS cameras.
The time intervals of high-flaring background contamination were identified and excluded by examining the light curves in the 10--12 keV energy range.
The total cleaned exposure times for the PN and MOS cameras are 28 and 33 ks, respectively, in 2009, and are 12 and 20 ks in 2022.
We extracted the source+background spectra from a circular region centred on the target with a 32$\arcsec$ radius, and the background spectra from an adjacent source-free circular region with a larger radius.
We combined the spectra taken from PN and two MOS cameras using the task {\it epicspeccombine}, during which the background spectra and the response matrices were also combined.
The combined EPIC spectra were then grouped so that there are at least 20 net counts in each energy bin to ensure a $\chi^2$ statistic.
We also extracted the photometry of F01004-2237 on the OM images using the task {\it omichain}, and the results are listed in Table~\ref{tab:uvotdata}.

\subsection{Optical spectroscopic observations} \label{sec:2.3}

We made follow-up optical spectroscopic observations of F01004-2237 using Lijiang 2.4-meter telescope (LJT) on December 20, 2021 and January 3, 2022, Southern Africa Large Telescope (SALT) on December 30, 2021, Magellan telescope on January 2, 2022, Gemini South telescope on July 3, 2022 (Proposal ID GS-2022A-DD-104), and Palomar 200-inch Hale telescope (P200) on September 3, 2022.

\begin{table}
\small
\centering
\caption{Summary of optical spectral observations}
\begin{tabular}{cccccccccc}
\hline
\hline
Instrument & obs-date & $\lambda$ range  & slit & SNR  & $R$ \\
(1)        & (2)      & (3)              & (4)  & (5) & (6) \\
\hline
VLT/XShooter    & 2018/08    & 2680--22170  & 1.2 & 65  & 6500 \\
LJT/YFOSC       & 2021/12/20 & 3040--6840   & 2.0 & 9.4 & 600 \\
SALT/RSS        & 2021/12/30 & 3990--6620   & 1.5 & 39  & 900 \\
Magellan/LDSS-3 & 2022/01/02 & 3380--8940   & 1.0 & 48  & 860 \\
LJT/YFOSC       & 2022/01/03 & 3750--6850   & 2.0 & 14  & 600 \\
Gemini-S/GMOS   & 2022/07/03 & 4350--8230   & 1.5 & 16  & 750 \\
P200/DoubleSpec & 2022/09/03 & 2870--9100   & 1.5 & 19  & 770 \\
\hline
\hline
\end{tabular}
\begin{tablenotes}
    \item (1): The name of the telescope and the spectrometric instrument.
    (2): The date of observation.
    (3): The wavelength coverage in the rest-frame in unit of \AA.
    (4): The width of slit in unit of arcsec.
    (5): The signal-to-noise ratio per \AA\ at 6000 \AA.
    (6): The spectral resolution at 6000 \AA.
\end{tablenotes}
\label{tab1}
\end{table}

\begin{figure}
\centering
  \includegraphics[scale=0.7]{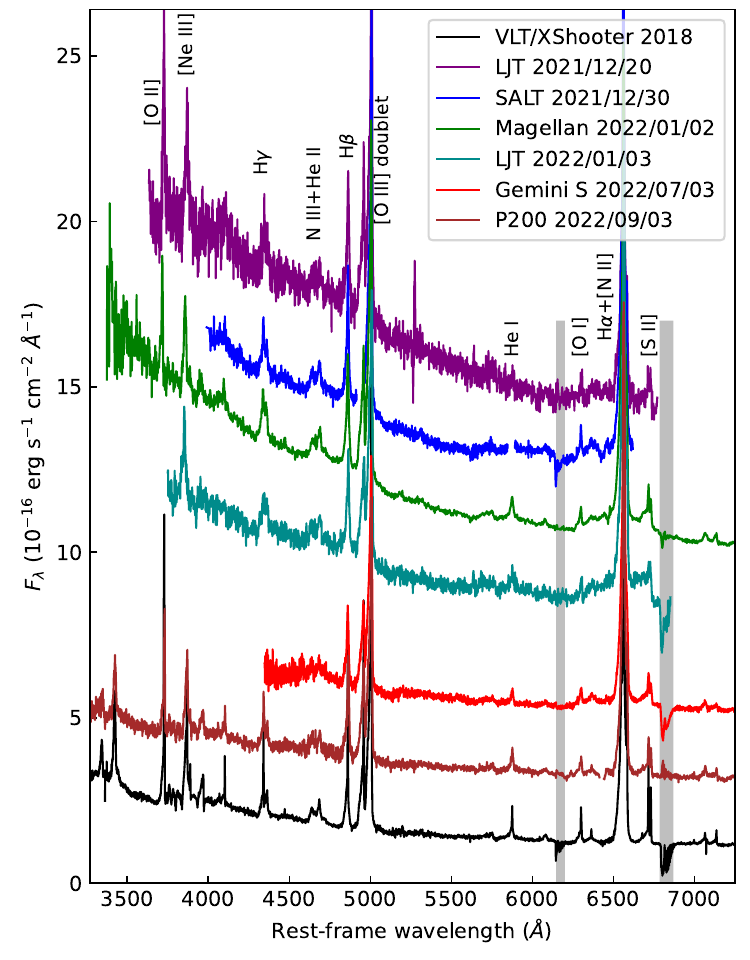}
  \caption{
  A collection of the optical spectra used in this work.
  The spectra are shifted vertically by different constants for clarity.
  The regions affected by telluric absorption are labeled using grey shades.
  }
\label{fig:show_optspec}
\end{figure}

For the SALT observation, we used the Robert Stobie Spectrograph \citep[RSS,][]{Burgh2003} with the PG0900 grating, resulting in a mean resolving power of $R \sim 800-1200$.
The exposure time is 1000 seconds.
To reduce the spectra, we used the PyRAF-based PySALT package\footnote{https://astronomers.salt.ac.za/software/} \citep{Crawford2010}, which includes corrections for gain and cross-talk and bias subtraction.
We extracted the science spectrum using standard IRAF tasks, including wavelength calibration, background subtraction, and 1D spectra extraction.
We obtained flux calibration using observations of standard stars during twilight.
For the Magellan observation, we used the Low Dispersion Survey Spectrograph (LDSS-3).
The total exposure time is 900 seconds.
We reduced the spectra using IRAF following the standard routine.
For the Gemini South observation, we used the GEMINI Multiple Object Spectrographs \citep[GMOS,][]{Davies1997} with an R400 grating, centring at 655 nm, resulting in a continuous wavelength coverage of 4860--9200 \AA.
Three exposures with 600 seconds exposure each were taken.
Following the standard routine, we reduced the data using Pyraf, during which the flux calibration was made using a standard star spectrum taken on the same night.
For the P200 observation, we used the Double Spectrograph (DBSP) with a D55 dichroic, a 600/4000 grating for the blue side, and a 316/7500 grating for the red side, resulting in a continuous wavelength coverage of 3200--10200 \AA.
The total exposure time is 700 seconds.
We reduced the data using Pyraf following the standard routine.

To model the starlight and identify narrow emission lines (NELs), which are crucial for analyzing the flare's spectra, we obtained the pre-flare VLT/XShooter spectrum observed in August 2018 \citep[see details in][]{Tadhunter2021}.
We downloaded the two-dimensional spectra from the two exposures from ESO database\footnote{http://archive.eso.org/wdb/wdb/adp/phase3\_spectral/form }, which had been processed by the XShooter pipeline, and then combined them to remove cosmic rays, and finally extracted one-dimensional spectra with an aperture of 1.8".
We recalibrated the fluxes of all the spectra using quasi-simultaneous photometric data.
All the spectra used in this work are shown in Fig.~\ref{fig:show_optspec}.

\subsection{UV spectroscopic observations} \label{sec:2.4}

\begin{figure}
\centering
  \includegraphics[scale=0.7]{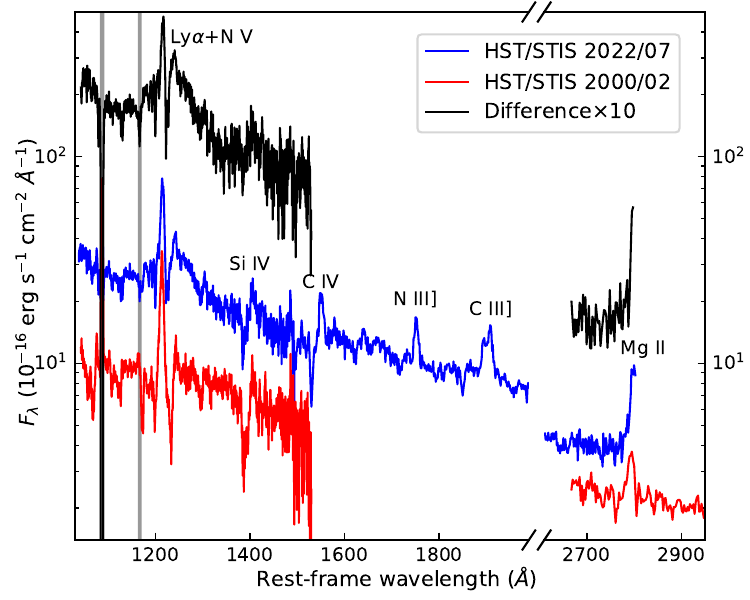}
  \caption{
  The \textit{HST}/STIS spectra taken in 2000 (red) and 2022 (blue), and the difference spectrum between them (black, multiplied by 10 for clarity).
  The spectrum in 2000--2600 \AA\ shows no emission feature and is not presented.
  We mark the regions affected by galactic emission lines (Ly$\alpha$ and O I $\lambda$1302) with grey shades.
  }
\label{fig:show_uvspec}
\end{figure}

We triggered a follow-up UV spectroscopic observation in July 2022 with Space Telescope Imaging Spectrograph (STIS) mounted on Hubble Space Telescope (\textit{HST}, proposal ID 16943).
The observations using G140L and G230L gratings were made on July 30 and July 14 with exposure times of 3766 and 5947 seconds, respectively.
A slit with an aperture of 52$\times$0.2 arcsec$^2$ was used and positioned through the galaxy centre.
We downloaded the data from the Mikulski Archive for Space Telescopes\footnote{https://archive.stsci.edu } (MAST).
We obtained the two-dimensional spectra that the STIS pipeline had processed.
We then combined the exposures to remove cosmic rays and extracted one-dimensional spectra with an aperture of 0.2$\arcsec$.
As a result, the spectrum is from a region in the galaxy centre with an area of 0.2$\times$0.2 arcsec$^2$, taking into account the slit width.

To obtain the host galaxy's contribution to UV emission, we collected an archival \textit{HST}/STIS spectrum (Proposal ID 8190) observed on February 9, 2000 \citep[see details in][]{Farrah2005}.
Observations with G140L, G430L and G750L gratings were taken.
A 52$\times$0.2 slit was also used and positioned through the galaxy centre.
With a similar procedure, we obtained a pre-flare spectrum, which is also from a 0.2$\times$0.2 arcsec$^2$ region in the galaxy centre.
We made subtraction with the two G140L spectra to obtain a difference spectrum covering a wavelength range of 1165 to 1710 \AA.
All the STIS spectra used in this work and the difference spectrum are shown in Fig.~\ref{fig:show_uvspec}.
Because the aperture sizes of the two spectra are the same and the flux calibration of STIS is stable, we adopted the difference spectrum as emission from the 2021 flare.

\section{Data analysis}

\subsection{Position of the 2021 flare} \label{sec:3.1}

We checked the position of the 2021 flare using \textit{Gaia} data with the best positional accuracy.
According to the \textit{Gaia} alert, when the 2021 flare was detected, the coordinate of the target was RA = 01h 02m 49.990s and Dec = -22d 21m 57.24s (J2000).
The coordinate before the flare occurred was RA = 01h 02m 49.991s and Dec = -22d 21m 57.27s, which was obtained from the source catalogue of \textit{Gaia} data release 2 \citep[DR2,][]{GaiaDR2}.
The offset between the two positions is 33 mas.
\cite{Hodgkin2021} found that the offset between \textit{Gaia} alerts and their \textit{Gaia} DR2 counterparts obeys a Rayleigh function with an average value of 55 mas.
Thus, we found no evidence that the flare's position deviates from the centre of the galaxy.
Using the Rayleigh function above, we can limit the offset of the two positions to be within 160 mas (99.73\% probability).

The constraint to the position of the \textit{Gaia} alert is not equivalent to the constraint to the flare's position.
This is because the \textit{Gaia} position is the average position based on the brightness distribution.
When an off-centre flare occurs, the average position should be between the flare's position and the centre of the galaxy.
The alerting magnitude of Gaia22ahd is 0.8 magnitudes brighter than historical values, which means that the flare is 1.1 times brighter than the host galaxy.
A simulation shows that the offset of a hypothetical off-centre flare should be 1.9 times the offset of the average position.
Thus, we can limit the flare's offset relative to the galaxy centre to be within 300 mas.

We also measured the flare's position using the LCOGT $g$-band data.
We measured the positions on the reference-subtracted images with high signal-to-noise ratios.
We then calculated an averaged position of RA = 01h 02m 49.987s and Dec = -22d 21m 57.21s, which has an 82 mas offset from the \textit{Gaia} position.
We estimated a positional error of $\sim$80 mas from the scatter of the positions.
Thus, the analysis of LCOGT images yields no evidence of offset between the flare's position and the galaxy centre and limits the offset to be within 240 mas.

The results using \textit{Gaia} and LCOGT data are consistent.
Taking them together, we can limit the offset to be within $\sim$0.2$\arcsec$, corresponding to a physical size of 440 pc.

\subsection{UV/Optical light curves} \label{sec:3.2}

\begin{figure}
\centering
  \includegraphics[scale=0.7]{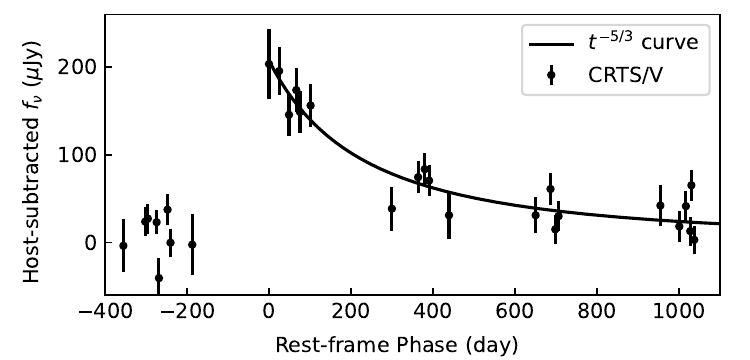}
  \caption{
  The host-subtracted LC in CRTS/$V$ band of the 2010 flare and the best-fitting $t^{-5/3}$ curve.
  }
\label{fig:lc_f2010}
\end{figure}

\begin{figure*}
\sidecaption
  \includegraphics[width=12cm]{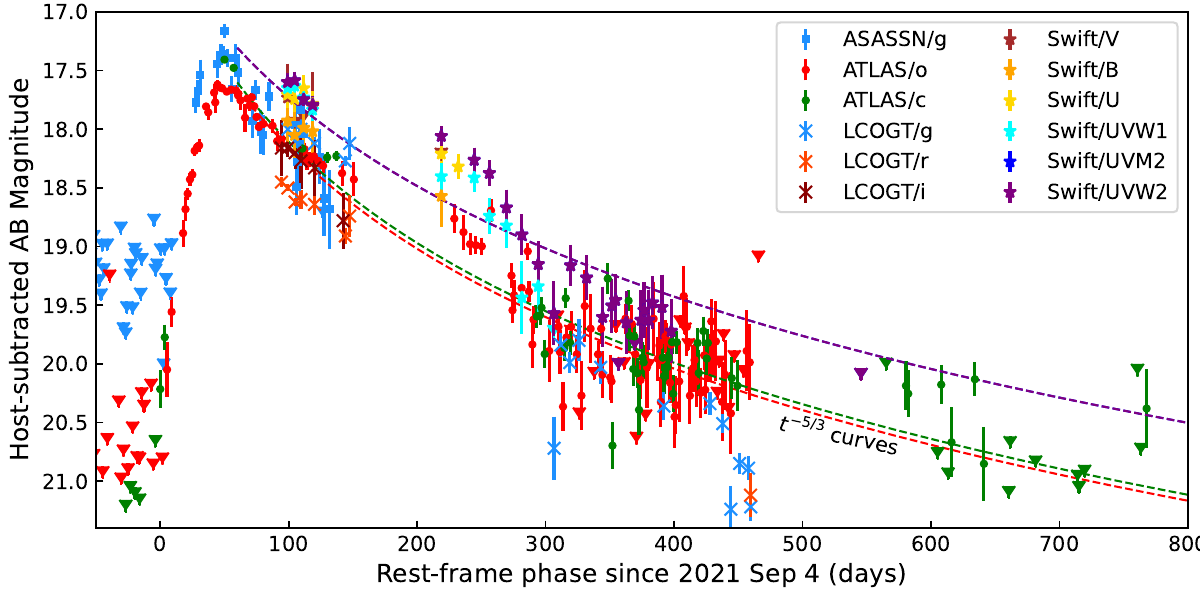}
  \caption{
  Host-subtracted LCs of the 2021 flare.
  The observation time was converted to phase relative to the discovery date in the rest frame.
  For clarity, we did not show the upper limits from ASASSN/$g$ observations after $+$200 days or those from ATLAS/$c$ observations after $+$500 days, and binned the LCOGT data weekly.
  We show the $t^{-5/3}$ curves that fit the data in UVW2, $c$, and $o$ bands after $+$60 days in dashed lines for comparison.
  }
\label{fig:lc_f2021}
\end{figure*}

In Fig.~\ref{fig:lc_opt}, we display the long-term variability of F01004-2237 with CRTS/$V$, ATLAS/$o$, ASASSN/$g$ and \textit{Gaia}/$G$ LCs, which clearly show the two flares.
The CRTS/$V$ LC shows that the galaxy's emission remained quiescent until January 23, 2010 (MJD$=$55219.5) with $V=17.54$, and rose by 0.5 magnitude to $V=17.04$ on September 17, 2010 (MJD$=$55456.4), and then slowly declined over the next three months, and then dropped to $V\sim17.3$ in 2011, and finally fading gradually.
With this LC, we inferred that the flare peaked between January and September 2010 (MJD$=55338\pm118$).
We fit the LC with a $t^{-5/3}$ curve expressed as:
\begin{equation}
f(t) = f_{\rm peak} \left( 1+ \frac{t-t_{\rm peak}}{t_0} \right)^{-5/3},
\end{equation}
where $t_{\rm peak}$ is the peak time and is set to be in the range of MJD$=55338\pm118$, and $f_{\rm peak}$ is the peak flux.
As can be seen in Fig.~\ref{fig:lc_f2010}, the curve well fit the data.
Using this curve, we estimated a peak $V$-band magnitude of $17.4\pm0.5$.
The \textit{Gaia}/$G$ LC shows a long-term fading trend from 2014 to 2018, so the 2010 flare might have lasted seven years.
The galaxy's emission went quiescent again from 2019 to August 2021, during which no variation was detected in any band.
Using data in this period, we derived the quiescent levels before the 2021 flare that $o=17.59$, $c=17.87$, and $g=17.81$.

In Fig.\ref{fig:lc_f2021}, we present the LCs of the 2021 flare after subtracting quiescent fluxes.
For the six \textit{Swift}/UVOT bands, we adopted the fluxes observed in May 2024 as the quiescent fluxes, because they are close to those observed by \textit{XMM-Newton}/OM in 2009 (the difference could be explained as statistical fluctuation, difference in filters' response, and variation of possible weak underlying AGN).
The first $>3\sigma$ detection was on September 4, 2021 (MJD=59461.6, the discovery date hereafter) in the ATLAS/$c$ band with $c=20.22\pm0.16$.
Then, the flare rapidly rose, reaching a peak about 50 days after discovery.
We fit the $o$-band LC near the peak with polynomial and obtained a $t_{\rm peak}$ of MJD$=59519.8\pm0.7$, and a rising time $t_{\rm rise}$ of $52$ days in the rest frame.
At $t_{\rm peak}$, the flare has $o=17.64\pm0.02$, $c=17.42\pm0.02$, and $g=17.36\pm0.07$.
Here, the $c$-band and $g$-band magnitudes were calculated by shifting the best-fitting $o$-band LC to match the data points in these bands.
These apparent magnitudes correspond to $M_g=-21.16$ and $M_V=-21.05$ after K correction.
The flare declined slowly after the peak in all bands, although some fluctuations could be seen.
After $+$500 days, the flare was only barely detected in the $c$ band and was swamped by host galaxy emission at other bands.
After $+$650 days, the flare dropped to an undetectable level with $c>21$.

\begin{figure}
\centering
  \includegraphics[scale=0.7]{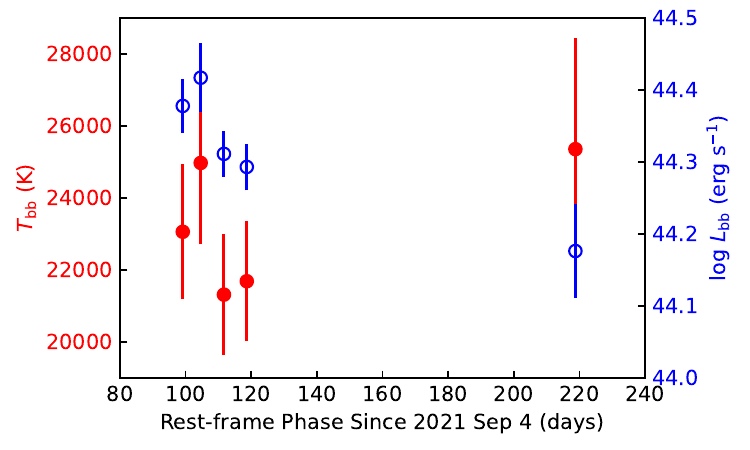}
  \caption{
  The time variation of the $T_{\rm BB}$ (red filled) and $L_{\rm BB}$ (blue empty).
  }
\label{fig:tbb_lbb}
\end{figure}

The LC that rises rapidly in 50--60 days, peaks with absolute magnitudes of $\sim-21$, and fades in years is reminiscent of superluminous supernovae (SLSNe) and TDEs.
To distinguish between the two possible origins, we measured the variation of the photospherical temperature.
We fit the SED obtained from the 5 \textit{Swift}/UVOT observations with all six filters used with blackbody curves.
We showed the $T_{\rm BB}$ and $L_{\rm BB}$ in Fig.~\ref{fig:tbb_lbb}.
The temperature keeps nearly constant as the luminosity declines and is still $>20,000$ K at a phase of $+220$ day.
Such a late-time high temperature is inconsistent with SNe and is consistent with TDEs \citep[e.g.,][]{vanVelzen2020,Zabludoff2021}.
In addition, as can be seen in Fig.~\ref{fig:lc_f2021}, the LCs in the dropping phase ($>$60 days) are roughly in line with $t^{-5/3}$ curves, also consistent with TDEs.

\begin{figure}
\centering
  \includegraphics[scale=0.7]{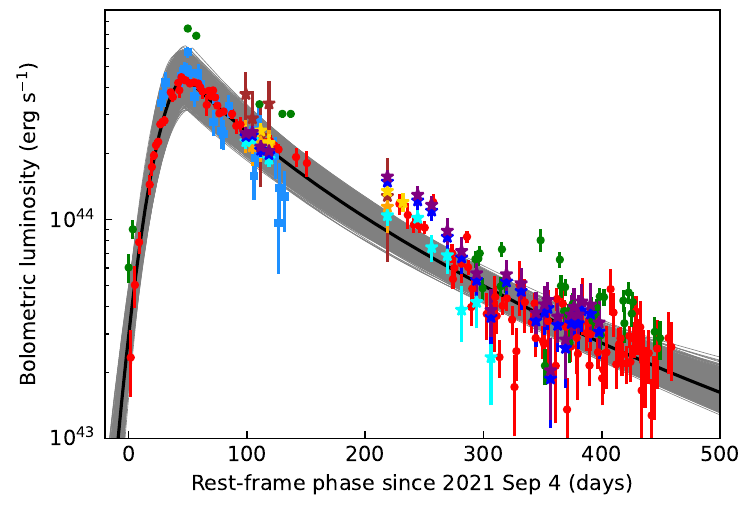}
  \caption{
  The MCMC realizations of bolometric LCs (grey) and their median LC (black).
  We show the observed data (colours the same as in Fig.~\ref{fig:lc_f2021}) after bolometric corrections for comparison.
  The data in the ATLAS/$c$ band (rest-frame 4840 \AA) show excess relative to the models, possibly due to broad He II and H$\beta$ emission lines (see section~\ref{sec:3.3} for details).
  }
\label{fig:tdemodel}
\end{figure}

To further examine the similarity between the flare and TDEs and measure the peak luminosity and the energy budget, we simultaneously fit the multi-band LC data with the TDE model of \cite{vanVelzen2021}.
The model assumes blackbody spectra with luminosity and temperature vary with time as:
\begin{equation}
L(t,\nu) = L_{\rm peak} \frac{\pi B_\nu\left( T(t) \right)}{\sigma_{\rm SB}T^4(t)}
   \times \left\{ \begin{array}{lr}
      { e^{-(t-t_{\rm peak})^2}/2\sigma^2 \ \ \ \ \ \ \ \ \ \ \   t<=t_{\rm peak}} \\
      { [(t-t_{\rm peak}+t_0)/t_0]^p  \ \ \     t>t_{\rm peak} }
    \end{array} \right. \\
\end{equation}
where $t_{\rm peak}$ and $L_{\rm peak}$ are the peak time and peak bolometric luminosity, $\sigma$ describes the how the pre-peak luminosity rises, and $t_0$ and $p$ describes the how the post-peak luminosity declines.
And the temperature $T(t)$ is assumed as:
\begin{equation}
T(t) = T_0 + {\rm d}T/{\rm d}t \times (t-t_{\rm peak}),
\end{equation}
We fit the data using the Markov Chain Monte Carlo (MCMC) code of \textit{emcee}, and set the priors following \cite{vanVelzen2021}.
The MCMC realizations of bolometric LCs are shown in Fig.~\ref{fig:tdemodel}, and the parameters are listed in Table~\ref{tab:tdelcpar}.
The peak time obtained using LCs in all bands agrees with the $o$-band peak time.
The peak bolometric luminosity is $4.4\pm0.4\times10^{44}$ erg s$^{-1}$, and the energy released up to $+$700 day reaches $5.2\pm0.4\times10^{51}$ erg.
We compared the parameters with those of the 17 ZTF TDEs of \cite{vanVelzen2021}, and found that most of the parameters of the 2021 flare fall within the range of typical values of TDEs, and the peak luminosity and the rise time are close to the maximum value of TDEs.

The UV LCs show a bump in $+220$ to $+260$ days on the basis of a power-law decline (Figure~\ref{fig:lc_f2021} and Figure~\ref{fig:tdemodel}).
Such late-time bumps, appearing as re-brightening or flattening on the LCs, are also seen in some TDEs and candidates, such as ASASSN-15lh \citep{Leloudas2016,Brown2016} and some recent TDEs \citep{Yao2023}.
Possible explanations include a transition from super-Eddington to sub-Eddington fallbacks \citep{Strubbe2009}, tidal disruptions of double stars \citep{Mandel2015}, and delayed formation of accretion disk \citep{Liu2022}.
Unfortunately, there are not enough observational data to test these possible explanations.

\begin{table}
\centering
\caption{TDE model parameters and typical values}
\begin{tabular}{ccc}
\hline
\hline
Parameters & 2021 flare & ZTF TDEs \\
\hline
log $L_{\rm peak}$ (erg s$^{-1}$)  & $44.65\pm0.04$ & 3.14 $\sim$ 44.62 \\
log $T$ (K)                        & $4.38\pm0.02$  & 4.07 $\sim$ 4.60 \\
$\frac{{\rm d}T}{{\rm d}t}$ ($10^2$ K day$^{-1}$) & $-0.15\pm0.03$ & $-0.40$ $\sim$ 1.94 \\
$t_{\rm peak}$ (MJD)               & $59516\pm3$ & - \\
log $\sigma$ (day)                 & $1.32\pm0.03$  & 0.1 $\sim$ 1.3 \\
$p$                                & $-3.6\pm0.5$   & $-4.0$ $\sim$ $-0.5$ \\
log $t_0$ (day)                    & $2.48\pm0.08$ & 1.23 $\sim$ 2.69 \\
\hline
\end{tabular}
\begin{tablenotes}
   \item
\end{tablenotes}
\label{tab:tdelcpar}
\end{table}

\subsection{Optical spectroscopy}  \label{sec:3.3}

\begin{figure}
\centering
  \includegraphics[scale=0.82]{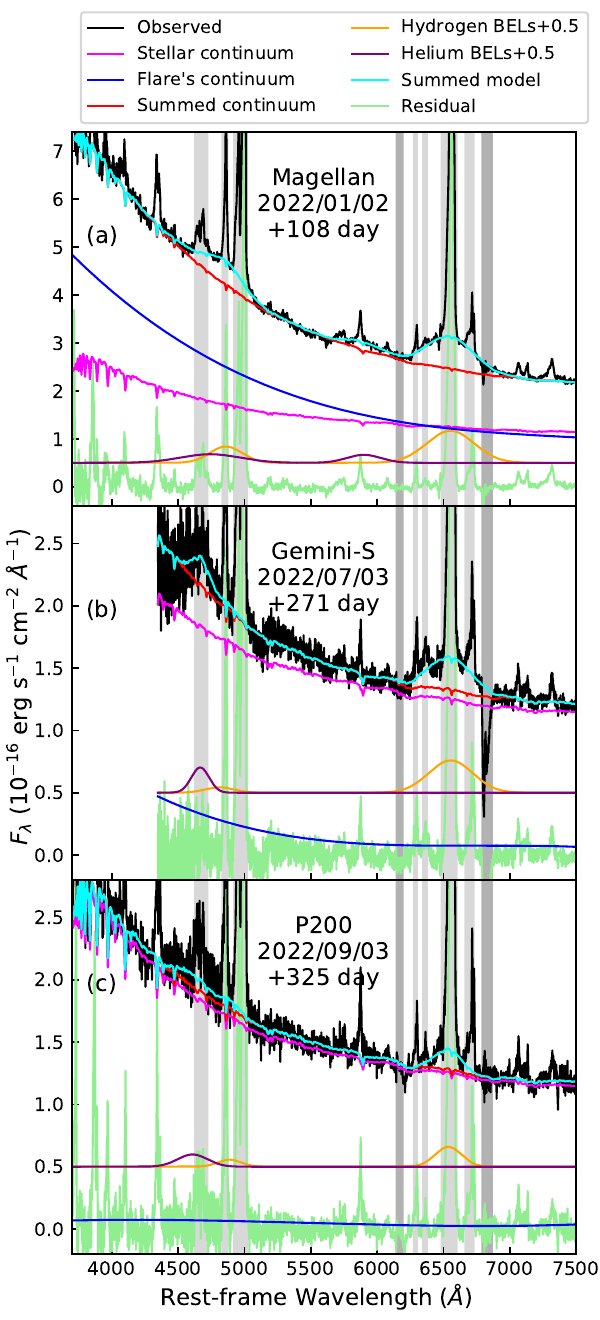}
  \caption{
  The spectra of the 2021 flare (black) and the best-fitting models (the MCMC realizations with the minimum $\chi^2$) assuming that the flare's continuum is a three-order polynomial.
  The meaning of the different colors are: magenta, the stellar continuum; blue, the flare's continuum; red, the sum of the two continuum components; orange, the hydrogen (H$\alpha$ and H$\beta$) VBELs; purple, the helium (He II $\lambda$4686 and He I $\lambda$5876) VBELs; cyan, the sum of all components.
  The wavelength ranges affected by NELs (light grey shades) and telluric absorption lines (dark grey shades) were not included in the fitting.
  On this figure, however, we only marked those affect H$\alpha$, H$\beta$ and He II VBELs for clarity.
  }
\label{fig:belfit}
\end{figure}

There are four spectra taken between $+$95 and $+$109 days (Fig.~\ref{fig:show_optspec}).
We first investigated the spectrum taken by Magellan on $+$108 day, which has the best data quality and the broadest wavelength coverage.
We modelled the spectrum with the sum of a stellar model and a third-order polynomial, the latter representing the flare's continuum emission, by masking the wavelength ranges affected by narrow emission lines from the host galaxy and telluric absorption bands.
We describe the construction of the stellar model and the identification of narrow emission lines in detail in Appendix~\ref{App:optspec}.
We fit using MCMC code \textit{emcee} assuming flat priors for all the parameters.
Although this model reproduces most parts of the spectrum, it leaves significant residuals in three wavelength ranges around 6550 \AA, 4830 \AA, and 5890 \AA, implying the presence of emission features (Fig.~\ref{fig:belfit}(a)).
We identified the 6550 \AA\ feature as H$\alpha$ and fitted it by adding a single Gaussian component to the model.
It is very broad with a Full Width at Half Maximum (FWHM) of $19200\pm400$ km s$^{-1}$ and is slightly blueshifted with a velocity of $550\pm150$ km s$^{-1}$.
We fit the 4830 \AA\ feature by adding another single Gaussian and found that the central wavelength is $4830\pm8$ \AA, the FWHM is $22400\pm1200$ km s$^{-1}$, and the flux is $0.62\pm0.03$ times that of H$\alpha$.
If identified as H$\beta$, the width and shifted velocity differ a lot from those of H$\alpha$, and the ratio of H$\alpha$/H$\beta$ of $1.62\pm0.08$ is less than the Case B value of 2.70\footnote{Here we adopted the value in $T_e=20,000$ K and $n_e=10^{10}$ cm$^{-3}$ \citep{Storey1995}.}.
These are not reasonable; hence, we considered that the 4830 \AA\ feature is a mixing of H$\beta$ and He II $\lambda$4686 emission lines.
We deblended them by adding two Gaussians for H$\beta$ and He II each and requiring that the FWHM and shifted velocity of H$\beta$ to be within $\pm$3,000 km s$^{-1}$ of those of H$\alpha$, following \cite{Charalampopoulos2022}.
We also added another single Gaussian for the 5890 \AA\ feature, identified as He I $\lambda$5876.
The model fits the data well, as seen in Fig.~\ref{fig:belfit}(a), and the parameters are listed in Table~\ref{tab:par_vbel}.
The inferred H$\alpha$/H$\beta$ ratio of $3.5^{+2.4}_{-0.6}$ is consistent with the Case B value.
The He II FWHM is $25100\pm3800$ km s$^{-1}$ and the He II/H$\alpha$ ratio is $0.35\pm0.08$.

To explore how much the assumption of the flare's continuum model would affect the analyses of BELs, we tried two other models, double blackbody and double power-law (they fit much better than either single blackbody or single power-law).
Through analyses similar to those described in the previous paragraph, we found that all four lines are significantly detected regardless of which model is used.
The best-fitting He II FWHM is $\sim$60,000 km s$^{-1}$.
We considered this value unreasonable because such broad He II emission lines have never been seen in TDEs or AGNs.
Therefore, we set the He II FWHM to be less than 25,000 km s$^{-1}$ (the value assuming polynomial continuum) in the MCMC.
The resultant BEL parameters are present in Table~\ref{tab:par_vbel}.
Combining the results of the three models, the H$\alpha$ FWHM is 16,000--19,000 km s$^{-1}$, the He II FWHM is $>21,000$ km s$^{-1}$, and He II/H$\alpha$ ratio is 0.3--0.7.

The three spectra observed between $+$95 and $+$109 days by SALT and LJT resemble the Magellan spectrum (Fig.~\ref{fig:show_optspec}), and VBELs can be seen visually.
We attempted to analyze them in the same way as we did for the Magellan spectrum.
For each of the three spectra, adding VBEL components significantly improved the goodness of fit.
This indicates that the VBEL persisted over this period of time.
However, the parameters of the VBELs have large uncertainties when different flare's continuum models are considered.
The reason may be the short spectral coverage, especially the absence of the red wing of H$\alpha$.
We will not use the results of these three spectra for further discussion.


The spectra observed by Gemini-S and P200 (Fig.~\ref{fig:belfit}(b) and ~\ref{fig:belfit}(c)) on $+$271 and $+$325 days still show BELs though the continuum flux decreases significantly.
We analyzed the two spectra similarly, except that we did not add a He I $\lambda$5876 component, which is not visible on the spectrum, and list the BEL parameters in Table~\ref{tab:par_vbel}.
H$\alpha$ and He II are still significantly detected, while the H$\beta$ is too weak to be detected.
In the Gemini-S spectrum, the width of H$\alpha$ BEL is 18,000--19,000 km s$^{-1}$, similar with that in the Magellan spectrum.
While in the P200 spectrum, the H$\alpha$ BEL FWHM is 7,000--11,000 km s$^{-1}$, much narrower.
The inferred He II/H$\alpha$ ratio is in the ranges of 0.3--0.6 and 0.9--2.3 for the two spectra, respectively.

Our analyses show that newly formed very broad emission lines (VBEL) of Balmer and He II $\lambda$4686 appear between $+$95 and $+$325 days after the 2021 flare.
The FWHMs of the VBELs exceed 7,000 km s$^{-1}$ consistently.
The width of H$\alpha$ VBEL can reach 16,000--19,000 km s$^{-1}$ at maximum, while that of He II can even reach $\gtrsim$21,000 km s$^{-1}$.

The VBELs in F01004-2237 seen 43--273 days after the optical maximum disagree with VBELs seen in SLSNe, which occur only near the optical peak \citep{Konyves-Toth2021}.
Combined with the analysis of late-time photospheric temperature in section 3.2, we ruled out the SLSNe interpretation for the 2021 flare.
In addition, the VBELs seen in F01004-2237 can be explained by TDE because such broad lines have been seen in other TDEs \citep{Charalampopoulos2022}.

\subsection{UV spectroscopy} \label{sec:3.4}

\begin{figure}
\centering
  \includegraphics[scale=0.7]{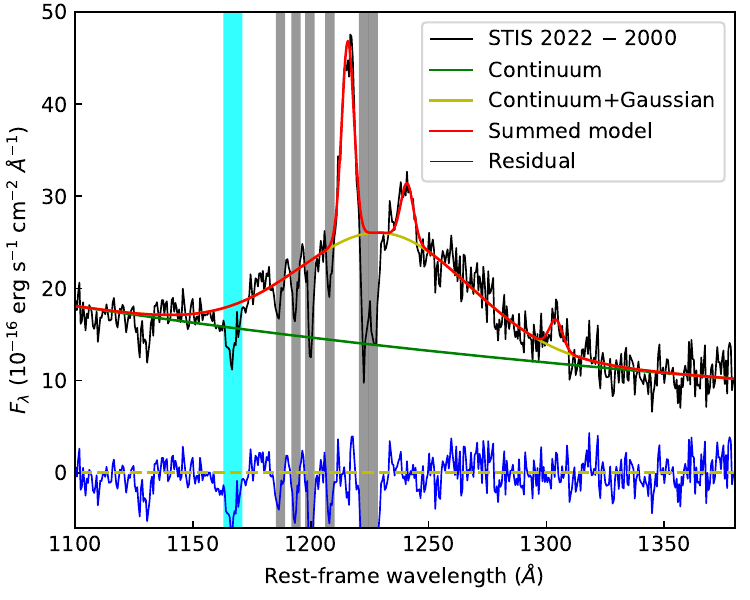}
  \caption{
  The UV spectrum of the 2021 flare and the best-fitting model, indicating the presence of Ly$\alpha$ VBEL.
  We labelled the regions affected by narrow absorption lines using grey shades and those affected by the Milky Way's emission lines using cyan shade.
  }
\label{fig:show_uvbel}
\end{figure}

We had obtained a UV spectrum in the rest-frame wavelength range of 1042--1530 \AA\ of the 2021 flare on July 30, 2022 ($+$294 day) using the \textit{HST}/STIS spectra with G140L grating taken in 2022 and 2000 (section 2.4).
The spectrum shows a broad emission feature around $\sim$1230 \AA\ on the continuum (Fig.~\ref{fig:show_uvbel}).
It also shows some narrower emission lines, located at 1216, 1241 and 1304 \AA, which were identified as Ly$\alpha$, N V $\lambda$1240 and O I $\lambda$1304, respectively.

We fit the spectrum with a model including a blackbody for the continuum, a Gaussian for the broad emission feature, and a Gaussian for each narrower emission line.
We masked the wavelength ranges affected by the narrow absorption lines (see details in Appendix~\ref{App:UVabs}) and the Milky Way's emission lines.
The best-fitting model is shown in Fig.~\ref{fig:show_uvbel}.
The broad emission feature has a central wavelength of $1231.5\pm0.8$ \AA, a FWHM of $20700\pm500$ km s$^{-1}$, and a luminosity of $4.5\pm0.1\times10^{42}$ erg s$^{-1}$.
If identified as Ly$\alpha$, it is redshifted by $3900\pm200$ km s$^{-1}$, and the FWHM is close to that of H$\alpha$ (18,000--19,000 km s$^{-1}$) in the GMOS spectrum with a time interval of 23 days.
The difference in the shifted velocity between Ly$\alpha$ and H$\alpha$ (around zero velocity) may be caused by intrinsic variation of the emission line region over the 23-day interval, the difference in conditions of blueshifted and redshifted emitting gas, or mixing of a possible N V $\lambda$1240 VBEL.
The flux ratio of Ly$\alpha$/H$\alpha$ is $10.4\pm0.4$, consistent with the Case B value of 13.0 considering the difference in UV and optical spectra apertures and a possible dust reddening.
Our analysis of the UV spectrum again confirms the presence of VBELs and suggests that UV and optical VBELs may have the same origin.

The narrower emission lines of Ly$\alpha$, N V $\lambda$1240 and O I $\lambda$1304 have a FWHM of $1500\pm80$ km s$^{-1}$, and is unshifted relative to the systematic redshift (velocity $20\pm30$ km s$^{-1}$).
This emission line system is also seen in C IV $\lambda$1550, He II $\lambda$1640, N III] $\lambda$1750, Si III] $\lambda$1892, C III] $\lambda$1907 and Mg II $\lambda$2800 in the G230L spectrum (Fig.~\ref{fig:show_uvspec}).
It may be the light echo from the nuclear ISM of the 2010 flare, the 2021 flare, or both of them.
A detailed analysis is beyond the scope of this paper and will be described in detail in the following works.

\subsection{X-ray variations and spectra} \label{sec:3.5}

\subsubsection{\textit{Swift}/XRT LC}

\begin{figure}
\centering
  \includegraphics[scale=0.7]{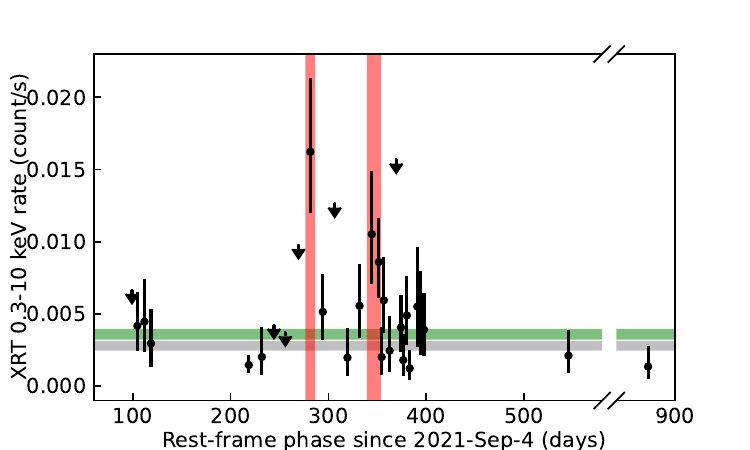}
  \caption{
  The XRT net count rates in 0.3--10 keV.
  The two X-ray flares are indicated by red shades.
  The averaged count rate (with 1$\sigma$ error) of all 29 observations and that of the other 26 are indicated by green and grey shades, respectively.
  }
\label{fig:xrt_rate}
\end{figure}

We searched for possible X-ray emission accompanied by the 2021 flare using the \textit{Swift}/XRT LC in 0.3--10 keV (Fig.~\ref{fig:xrt_rate}).
We found that the count rate in the $+$281 day observation is higher than that of other observations.
In addition, the rates in the two observations around $+$350 day also show weak excesses.
The count rate excesses indicate two X-ray flares, which were referred to as XF1 and XF2.
We calculated the significance levels of the two X-ray flares, that is, the probabilities that the count rate excesses are caused by statistical fluctuations under the assumption that the actual count rate is constant.

We calculated the weighted average count rate for all the observations.
The weight for the $i$th observation is set to be the effective exposure time $w_i=t_{\textrm{exp},i}$/$f_{c,i}$.
Thus, the weighted average rate is expressed as:
\begin{equation}
\bar{r} = \frac{\sum_i w_i r_i}{\sum_i w_i} = \frac{\sum_i N_{\textrm{tot},i} -  \sum_i N_{\textrm{bkg},i}}{\sum_i (t_{\textrm{exp},i}/f_{c,i})}
\end{equation}
We calculated the error using the Monte Carlo method, during which $\sum_i N_{\textrm{tot},i}$ and $\sum_i N_{\textrm{bkg},i}$ were realized assuming Poisson and Gaussian distributions, respectively.
The resultant rate is $3.49\pm0.36\times10^{-3}$ cts s$^{-1}$.

\begin{table*}
\centering
\caption{Significance of the two X-ray flares.}
\begin{tabular}{ccccccc}
\hline
\hline
    & phase (day)  & rate ($10^{-3}$ cts s$^{-1}$)  & $\sum N_{\textrm{tot}}$ & $E(\sum N_{\textrm{tot}})$ & $p$ (local) & $p$ (global) \\
\hline
XF1 & 281.5        & $16.23^{+5.11}_{-4.24}$ & 13 & $3.09\pm0.28$ & $1.48\times10^{-5}$ (4.33$\sigma$) & $4.28\times10^{-4}$ (3.52$\sigma$)\\
XF2 & 344.4--351.6 & $9.32\pm2.24$           & 19 & $7.66\pm0.70$ & $4.86\times10^{-4}$ (3.49$\sigma$) & $1.35\times10^{-2}$ (2.47$\sigma$)\\
\hline
\end{tabular}
\begin{tablenotes}
    \item
\end{tablenotes}
\label{tab:xflare}
\end{table*}

We estimated the probability of false detection of a flare as follows.
We first calculated the expected number of total photons in the source region during a flare as:
\begin{equation}
E\left( \sum_i N_{\textrm{tot},i} \right) = \sum_i N_{\textrm{bkg},i} + \bar{r}\times \sum_i (t_{\textrm{exp},i}/f_{c,i})
\end{equation}
The error of this expectation was calculated using the propagation of uncertainties.
We then generated 1000 realizations of $E(\sum_i N_{\textrm{tot},i})$ assuming a Gaussian distribution.
For the $j$th realization with an expectation of $E_j(\sum_i N_{\textrm{tot},i})$, we calculated the probability $p_j$ that $\sum_i N_{\textrm{tot},i}$ is greater than the observed value assuming Poisson distributions.
We finally averaged the 1000 $p_j$ to obtain a probability $p$.
We listed the results of two flares in Table~\ref{tab:xflare}, and found that both flares are detected by $>3\sigma$.
From the above calculations, we had obtained the local probability, which means the probability that statistical fluctuations cause a false excess signal in a given observation.
We were also concerned with the global probability, which means the probability of a false detection across all observations.
We calculated the global probability as $1-(1-p)^N$ for XF1, where $N=29$ is the total number of observations, and as $1-(1-p)^{N-1}$ for XF2, because XF2 was detected from two consecutive observations.
The final global probabilities listed in Table~\ref{tab:xflare} show that XF1 is significantly detected with 3.52$\sigma$, while XF2 is marginally detected with 2.47$\sigma$, and the hypothesis of no variation can be ruled out with a significance of 4.3$\sigma$.
After excluding the two X-ray flares, the average count rate for the remaining 26 observations is $2.73\pm0.33\times10^{-3}$ cts s$^{-1}$.
Then the XRT 0.3--10 keV count rate increased by factors of $6.0^{+2.1}_{-1.6}$ and $3.5\pm1.0$ during the two flares, respectively.

\subsubsection{Spectral analysis}

The X-ray emission of F01004-2237 shows intra-week variability in 2022, as seen from the XRT LC.
Such a short time scale can be explained as the X-ray emission associated with the 2021 flare or a possible underlying AGN.
To better distinguish between these two possibilities, we analyzed the X-ray spectra, including the new \textit{XMM-Newton}/EPIC spectrum observed in 2022 ($+$247 day) and the \textit{Swift}/XRT spectrum during XF1 ($+$281 day).
We also analyzed the pre-flare spectrum taken by \textit{XMM-Newton}/EPIC in 2009 for reference.
We fit all the spectra with \texttt{XSpec} (version 12.13).
All the errors reported in this subsection are at 90\% confidence.

\begin{figure}
\centering
  \includegraphics[scale=0.7]{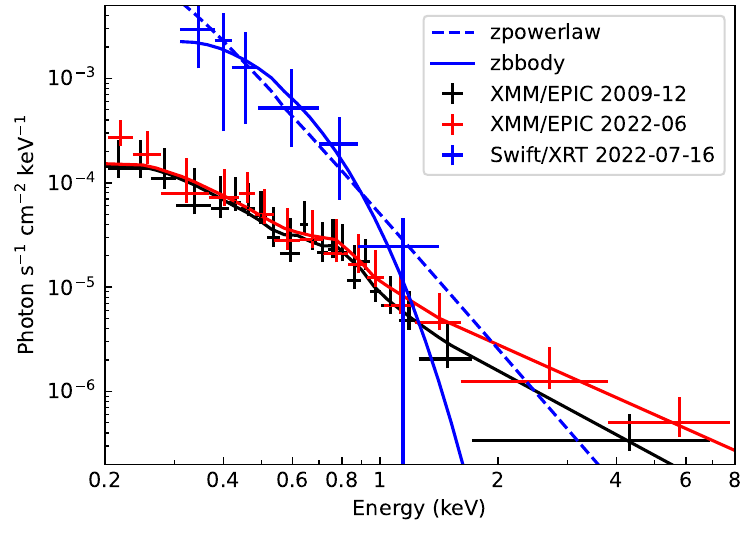}
  \caption{
  The \textit{XMM-Newton}/EPIC spectra taken in 2009 (black) and in 2022 ($+247$ day, red), and the \textit{Swift}/XRT spectrum taken in July 16, 2022 ($+281$ day, blue).
  We show the best-fitting models with corresponding colours.
  For the \textit{Swift} spectrum, we show the power-law and blackbody models in dashed and solid lines, respectively.
  }
\label{fig:xspec}
\end{figure}

\cite{Nardini2011} had modeled the 2009 spectrum by the sum of a \texttt{MEKAL} component and a powerlaw component.
We fit the spectrum with the same data (black in Fig.~\ref{fig:xspec}), yielding a temperature of $kT=0.66\pm0.09$ keV for the \texttt{MEKAL} component and an index of $\Gamma=2.62^{+0.17}_{-0.15}$ for the power-law component, which is consistent with the values of \cite{Nardini2011} within the margin of error.
The parameters suggest a starburst origin for the X-ray emission in 2009.
The EPIC count rate in the 2022 observation increased by 3.4$\sigma$, or by a factor of $1.45\pm0.25$, compared with that in the 2009 observation.
This suggests an additional X-ray emission component in 2022.
To investigate the spectral shape of the additional component, we fit the 2022 spectrum with a model containing three components: the first two are the same as those that fit the 2009 spectrum with fixed parameters, while the third is an additional power-law with free parameters.
The model fits the data well with $\chi^2/\textrm{d.o.f.}$ of $71.2/73$ (red in Fig.~\ref{fig:xspec}).
The additional component has a power-law index of $1.16\pm0.62$, implying that it is rather hard.

The spectrum of the XF1 on $+$281 day is rather soft, and only photons below 1.5 keV were detected.
A powerlaw model with $\Gamma=4.35^{+1.44}_{-1.29}$ and a blackbody model with $kT=113^{+40}_{-28}$ eV can both fit the spectrum with $C$ values of 0.6 and 0.7, respectively, for $\textrm{d.o.f.}=4$.

\subsubsection{Origin of the X-ray emission}

The additional component in the 2022 \textit{XMM-Newton} spectrum is flat with a power-law index of $1.16\pm0.62$.
Such a flat spectrum is unlikely to be produced by a TDE because TDEs typically show soft and steep spectra \citep{Saxton2020}.
Although some TDEs have flat and hard power-law components in their X-ray spectra \citep[e.g., ASASSN-15oi,][]{Holoien2016_15oi}, these components are much weaker than the soft thermal components and do not conform to the observation of F01004-2237.
A possible underlying AGN can explain the flat spectrum because the spectral index roughly agrees with the typical AGN values of $\Gamma\sim1.9$ \citep{Piconcelli2005} within the margin of error.
The 2--10 keV luminosities in the 2009 and 2022 spectra are $1.1\pm0.3\times10^{42}$ and $2.7\pm0.8\times10^{41}$ erg s$^{-1}$, respectively.
Considering the starburst contribution in 2009, the AGN X-ray luminosity has changed by a factor of $>$4, which is typical for long-term X-ray variability of AGNs.
Thus, the difference between the two \textit{XMM-Newton} spectra can be interpreted as a long-term variation of AGN.
We will discuss in Section 4.1 whether such an AGN exists based on multi-band data.

We also considered the possibility that the additional component could be related to the 2010 flare.
Because the 2010 flare caused a [Fe VII] and [Fe X] coronal emission line echo \citep{Tadhunter2017,Tadhunter2021}, it must be X-ray bright, although there are no direct X-ray observations to prove this.
The X-ray emission of the 2010 flare should have faded over 12 years, and possible late-time emission should be soft and cannot explain the observed flat spectrum.
However, it is also possible that the X-ray peak luminosity of the 2010 flare was Compton scattered by surrounding gas and caused a Compton echo, as was proposed to explain the diffused X-ray emission in the galactic centre \citep{Koyama1996,Yu2011} and the X-ray afterglow of Swift J1644+57 \citep{Cheng2016}.
Compton echoes are expected to have relatively flat spectra, consistent with the observations.
A large amount of ionized gas at pc-scale is required to produce an echo lasting for a decade.
This condition is satisfied in F01004-2237 because the coronal emission line echo lasts for at least 8 years \citep{Tadhunter2021}.
Thus, we inferred that the additional component in June 2022 has two possible origins: the long-term variability of an AGN with X-ray luminosity of several $10^{41}$ erg s$^{-1}$, or a Compton echo from the X-ray bright 2010 flare.

The X-ray flares discovered from the XRT LC on around $+$280 and $+$350 days have variability time scales of no more than 2--3 weeks and, therefore, must come from the vicinity of the SMBH.
Its spectrum is too steep (powerlaw index $\Gamma=4.35^{+1.44}_{-1.29}$) to be emission of an AGN.
Some galaxies have been reported to be AGNs with super-soft X-ray emissions, such as GSN 069 \citep{Miniutti2013} and 2XMM J123103.2+110648 \citep{Lin2013}.
However, they were later reclassified as TDEs \citep{Shu2018,Lin2017} based on the long decay of their X-ray fluxes.
Therefore, it is more likely that a TDE causes the X-ray flares.
X-ray flares with less than 2--3 weeks time scales have been observed in other TDEs, such as AT 2019ehz and AT 2023lli \citep{vanVelzen2021,Huang2024}.
The flares were interpreted as a sudden decrease in the optical depth of the absorbing material enveloping the X-ray emitting region, possibly caused by the movement of discrete absorbing clouds.
This interpretation may also apply to the 2021 flare of F01004-2237, during which the X-ray emission from the nuclear region is obscured by the envelope for most of the time, and is exposed only during the two X-ray flares.

\subsection{Radio observations}

\begin{figure}
\centering
 \includegraphics[scale=0.52]{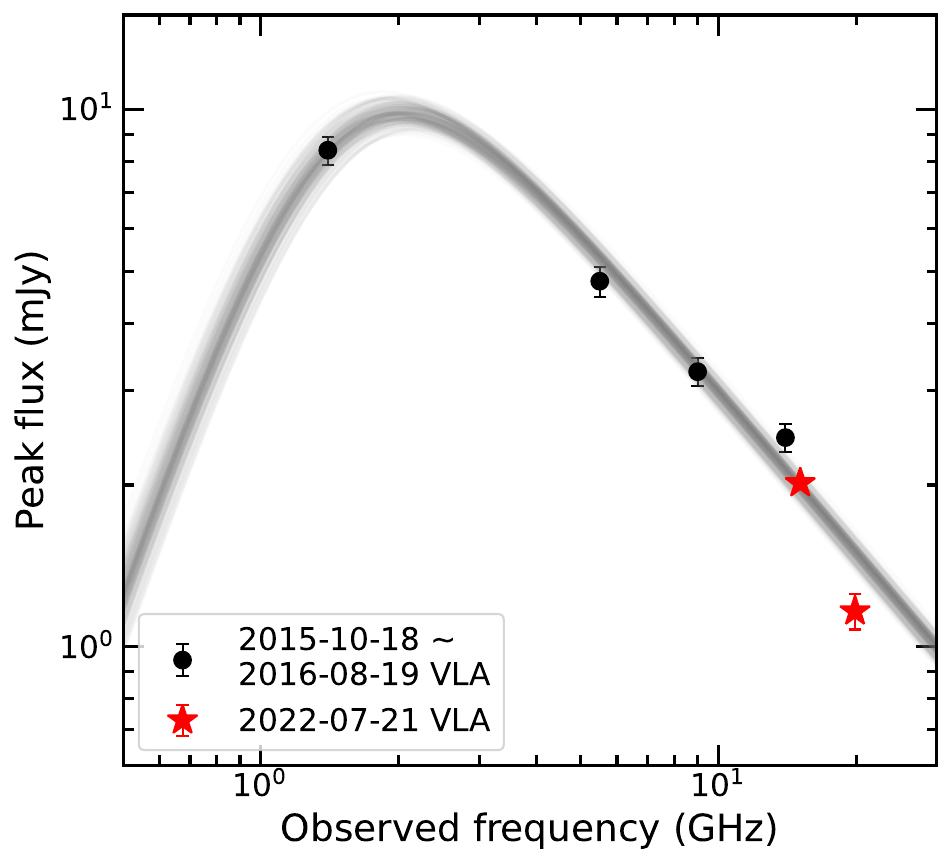}
  \caption{
  The radio SED constructed from the data taken from archival VLA observations covering frequency between 1.4 and 14 GHz \citep[black,][]{Hayashi2021}, and that from new VLA observations centered at $\sim$15 and $\sim$20 GHz (red).
  We show the MCMC realizations of the synthrotron emission model with gray lines.}
\label{fig:radio_sed}
\end{figure}

Prior to its 2021 flare, F01004-2237 was observed and detected by VLA at 1.4, 5.5, 9.0, and 14.0 GHz between 2015 Oct and 2016 Aug \citep{Hayashi2021}.
As shown in Fig.~\ref{fig:radio_sed}, the radio SED can be fitted with a synthrotron emission model using an MCMC fitting technique \citep[e.g.,][]{Goodwin2022, Zhang2024}.
In order to probe whether the radio emission is enhanced following the 2021 flare of F01004-2237, i.e., a nascent jet or outflow was launched, we proposed high frequency (Ku- and K-band) observations with VLA (Program ID 22A-507; PI, Shu) on 2022 July 21.
The data were reduced following standard procedures with the CASA package \citep{CASA2022}, and the source is clearly detected at both bands.
We used the IMFIT task in CASA to fit the radio emission component with a two-dimensional elliptical Gaussian model to determine the position, and the integrated and peak flux density.
A comparison of the integrated flux to the peak flux density indicates that the radio emission is compact, and no extended component is detected.
By plotting the integrated flux density in Fig.~\ref{fig:radio_sed}, we found no excess in the radio emission in comparison with the flux densities extrapolated from the best-fit synthrotron emission model.
Therefore, there is no obvious radio brightening since the 2021 optical flare, at least at the time of our VLA radio observations.

IRAS F01004-2237 was also observed by the Very Long Baseline Array (VLBA) on 2022 May 6 \citep{Hayashi2024} although the observation was not designed to study this flare.
A compact source was detected at 8.4 GHz on a 1 pc scale.
Its position is consistent with the galaxy center measured by Gaia with an offset of only 3 mas, and its peak flux is 0.6 mJy beam$^{-1}$.
The compact source could be related to the long-standing AGN, or it could be related to the optical flare.
With the available data, we cannot distinguish between these two possibilities.
New high-resolution radio observations in the future may detect changes in the flux or position of this compact source, thereby solving the mystery of its origin.

\section{The nature of the 2021 flare}

In this section, we discuss the nature of the 2021 flare.
We have ruled out a possibility of SLSNe with a nearly constant photospheric temperature of $\sim$22000 K from $+$99 to $+$219 day and VBELs with H$\alpha$ FWHMs up to 17,000--19,000 km s$^{-1}$ from $+$95 to $+$325 day.
We therefore considered two possible origins associated with SMBH accretion, including TDE and AGN flare.

\subsection{Constrains on pre-flare nuclear activity} \label{sec:4.1}

We checked the level of possible nuclear activity before the flare occurred.
This is not easy, as F01004-2237 shows vigorous starburst activity, which could overwhelm possible AGN signals.
To make matters worse, there are $\sim10^5$ Wolf-Rayet (WR) stars in the nucleus \citep{Armus1988,Farrah2005,Tadhunter2017}, producing some features similar to AGNs.

No broad emission lines (BELs) were detected in any of the optical spectra before the 2010 flare, including CTIO spectrum in 1985 \citep{Armus1988}, KPNO spectrum in 1995 \citep{Veilleux1999,Lipari2003}, \textit{HST}/STIS spectrum in 2000 \citep{Farrah2005}, and WHT spectrum in 2005 \citep{RodriguezZaurin2013}.
There is neither Pa$\alpha$ BEL in the NIR spectrum taken in 1995 \citep{Veilleux1997}.
The non-detection of BELs, even in the NIR band, cannot be explained by the obscuration of extremely thick dust.
This is because the vicinity of the SMBH must be seen directly as VBELs with FWHMs $>16,000$ km s$^{-1}$ were detected after the 2021 flare, with fluxes varying within months, and the ratio of Ly$\alpha$/H$\alpha\sim10$ indicates little dust reddening.
Therefore, the non-detection of BELs is because they are intrinsically weak.

In addition, the X-ray spectra before the 2010 flare can be explained by starburst alone, without needing a power-law AGN component \citep{Teng2010,Nardini2011}.
This cannot be explained by a Compton thick AGN, as proposed by \citep{Nardini2011}, because X-ray emission with a steep spectrum varying in weeks was detected after the 2021 flare, indicating little gas absorption in the line of sight.
Thus, we concluded that the nuclear activity must be weak before the 2010 flare.
We had measured a 2--10 keV luminosity of $2.7\pm0.8\times10^{41}$ erg s$^{-1}$ using the 2009 spectrum, consistent with that measured by \cite{Nardini2011}.
We adopted this luminosity as the upper limit of X-ray luminosity of AGN in 2009 and estimated an upper limit of bolometric luminosity of $\sim3\times10^{42}$ erg s$^{-1}$ using a correction factor of 12 from \cite{Hopkins2007}.


Some literature claimed to have found evidence for luminous AGN \citep[e.g.,][]{Veilleux2009,Yuan2010}.
We have re-examined their results and will discuss them in section 6.1.

\subsection{TDE versus AGN flare} \label{sec:4.2}

We discuss the two possible interpretations of the 2021 flare in terms of shapes of LC, UV/optical SED and luminosities, parameters of VBELs, and X-ray to UV/optical luminosity ratios.

TDEs typically have smooth LCs rising quickly and falling slowly, in line with the observation of the 2021 flare, which rose in $\sim$50 days and dropped by 2--3 magnitudes in one year.
However, the LCs of AGN flares, taking observations of changing-look AGNs as examples \citep[e.g.,][]{Ricci2023}, showed rich diversity without a uniform pattern.
Some AGN flares have LCs resembling those of TDEs, such as AT 2018dyk \citep{Frederick2019}.
Thus, the LCs support a TDE interpretation but do not exclude the possibility of AGN flare.

The UV/optical SEDs of the 2021 flare can be well described by blackbody curves with temperatures $\sim$22,000 K, consistent with optical TDEs \citep[e.g.,][]{vanVelzen2021,Hammerstein2023}.
Meanwhile, the UV-optical SEDs of AGNs are generally close to power-law curves.
However, it is challenging to distinguish a blackbody with a temperature of $\sim$22,000 K and a slightly extinct power-law, as they have little difference in observable wavelength ranges.
We tried to model the observed SEDs from \textit{Swift}/UVOT with power-law curves, yielding similarly good fits as blackbody curves with no statistically significant $\chi^2$ differences.
We did not make a similar distinction using spectral data considering the interference from observed VBELs, possible other VBELs from He and metal elements, and possible Balmer continuum and Fe II emission.
If assuming a SED similar to AGNs, we estimated a peak bolometric luminosity of $\sim9\times10^{44}$ erg s$^{-1}$ using the measured peak $\nu L_\nu$ at 4400 \AA\ of $9\times10^{43}$ erg s$^{-1}$ and adopting a bolometric correction of $\sim$10 from \cite{Hopkins2007}.
We had demonstrated in section 4.1 that the pre-flare nuclear activity, if present, had a bolometric luminosity of $<3\times10^{42}$ erg s$^{-1}$.
Then, the variation has an amplitude of more than two orders of magnitude, which is rare among AGN flares.
Thus, the UV/optical SED and luminosities support the TDE interpretation but do not exclude AGN flare.

\begin{figure}
\centering
  \includegraphics[scale=0.7]{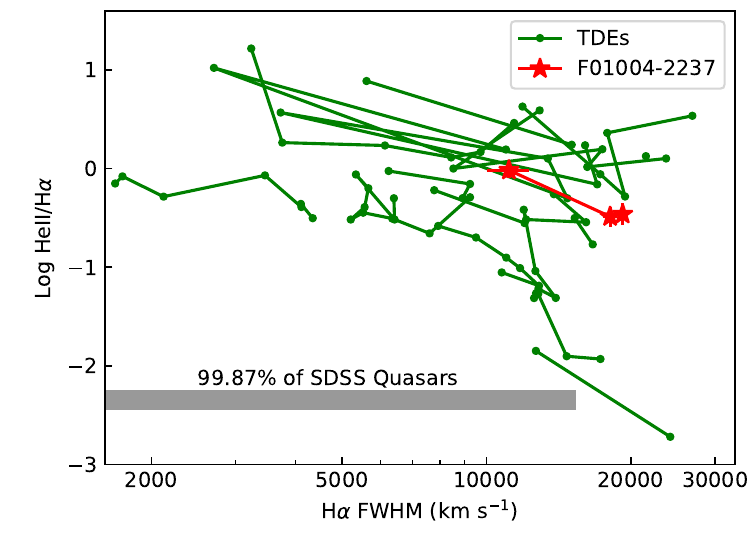}
  \caption{
  A comparison of the H$\alpha$ FWHM and He II/H$\alpha$ ratio of the 2021 flare with those of TDEs \citep{Charalampopoulos2022} and quasars.
  For the 2021 flare, we show the values assuming that the flare's continuum is a third-order polynomial, and the values are located in the TDE region in this figure.
  If the results by assuming other continuum model are adopted, the values will still be in this region.
  We present the range of H$\alpha$ FWHMs of 99.87\% of SDSS DR7 quasars from \cite{Shen2011} and the averaged He II/H$\alpha$ ratio of SDSS early data release quasars from \cite{VandenBerk2001}.
  }
\label{fig:bel_tde_agn}
\end{figure}

\begin{figure}
\centering
  \includegraphics[scale=0.7]{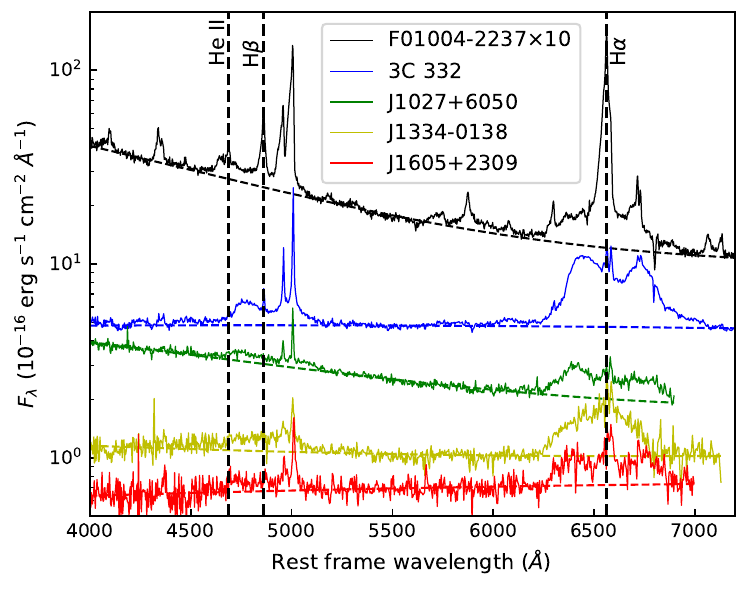}
  \caption{
  Comparison of the Magellan spectrum of F01004-2237 and the SDSS spectra of the four quasars with FWHM $>$16,000 km s$^{-1}$.
  Best-fitting polynomial continuum models are shown in dashed lines.
  }
\label{fig:quasar_vbel}
\end{figure}

The 2021 flare shows VBELs in H$\alpha$, H$\beta$, He II $\lambda$4686, He I $\lambda$5876 and Ly$\alpha$ between $+$105 and $+$325 days.
The FWHM of H$\alpha$ can reach 16,000--19,000 km s$^{-1}$, and that of He II can even reach $>$21,000 km s$^{-1}$, although the latter relies on the decomposition of He II with H$\beta$.
VBELs and strong He II emission distinguish TDEs from AGNs \citep{Zabludoff2021}.
As can be seen from Fig.~\ref{fig:bel_tde_agn}, the H$\alpha$ FWHMs and He II/H$\alpha$ ratios in F01004-2237 are consistent with those of TDEs \citep{Charalampopoulos2022}.
To understand the properties of BELs in AGNs, we selected 3119 SDSS DR7 quasars with a signal-to-noise ratio of H$\alpha$ BEL $>$10 using measurements by \cite{Shen2011}.
Among them, 99.87\% have H$\alpha$ FWHM $<$16,000 km s$^{-1}$.
\cite{Shen2011} did not measure the fluxes of He II, nor could we find any other literature that systematically investigated the He II fluxes in AGNs.
This is not surprising, as in most quasars and Seyfert 1 galaxies, He II BELs can not be detected because they are weak and overwhelmed in the Fe II bump.
\cite{VandenBerk2001} measured the He II and H$\alpha$ fluxes in the composite spectrum of SDSS quasars.
With their measurements, we estimated an average He II/H$\alpha$ ratio in quasars of 0.0045, much smaller than in F01004-2237.
We found four SDSS DR7 quasars (0.13\%) with H$\alpha$ with FWHM $>$16,000 km s$^{-1}$, and show their spectra in Fig.~\ref{fig:quasar_vbel}.
One of them is a well-studied quasar 3C 332.
It shows BELs with clear double-peak structures, which are thought to be emission from an accretion disk viewed from an inclined line of sight \citep{Halpern1990}.
Of the other three, J1027+6050 and J1605+2309 also show H$\alpha$ with a double-peak structure, and J1334-0138 has an H$\alpha$ emission line profile that deviates from a single Gaussian.
None of the four show clear He II BELs, as their weak emission features around 5000 \AA\ can be explained by H$\beta$.
All of these are inconsistent with VBELs in F01004-2237.
Therefore, the properties of VBELs strongly favour the TDE interpretation other than the AGN flare interpretation.

Most of the time after the 2021 flare, the X-ray emission is weak.
During two X-ray flares around $+$280 and $+$350 day, there are significant enhancements in X-ray emission lasting no more than weeks.
On one hand, weak and highly variable X-ray emission is consistent with TDEs.
Only 4 out of 17 ZTF TDEs in the sample of \cite{vanVelzen2021} show clear X-ray detection.
Their optical-to-X-ray ratios, expressed as ratios between UV/optical blackbody luminosity and 0.3--10 keV X-ray luminosity, are between 1 and 2000.
The same ratio during XF1 in F01004-2237 is 2--6, which is calculated assuming a power-law or a blackbody X-ray spectrum, and is consistent with X-ray bright TDEs.
On the other hand, X-ray emission is prevalent in AGNs.
The relative strength of X-ray emission in AGNs is described using $\alpha_{\rm OX}$, calculated with the fluxes at 2 keV and 2500 \AA\ \citep{Tananbaum1979}.
We computed $\alpha_{\rm OX}$ of the 2021 flare using \textit{XMM-Newton} data on $+$247 day and \textit{Swift}/XRT data on $+$281 day, and list the results in Table~\ref{tab:alphaox}.
The former was considered to be a lower limit because the X-ray emission on $+$247 day does not necessarily come from the 2021 flare.
The latter was calculated assuming a power-law or a blackbody X-ray spectrum, respectively.
Using the correlation between $\alpha_{\rm OX}$ and $L(\nu)_{2500}$ in AGNs \citep{Lusso2010}, the predicted $\alpha_{\rm OX}$ is only $\sim1.2$.
The observed values are larger than the predicted values after considering the scatter of the correlation of 0.15, indicating that the X-ray emission is weaker relative to the UV emission during the 2021 flare than in AGNs.
Therefore, the X-ray to UV/optical luminosity ratios favor the TDE interpretations.

\begin{table}
\centering
\caption{Measured $\alpha_{\rm OX}$ and prediction of AGN SED.}
\begin{tabular}{cccc}
\hline
\hline
phase & measured          & Log $L(\nu)_{2500}$      & predicted  \\
(day) & $\alpha_{\rm OX}$ & (erg s$^{-1}$ Hz$^{-1}$) & $\alpha_{\rm OX}$ \\
\hline
247.4                   & $>1.9$             & 28.76                  & 1.25\\
\hline
\multirow{2}{*}{281.5}  & $1.58\pm0.08$ (pl) & \multirow{2}{*}{28.41} & \multirow{2}{*}{1.20}\\
                        & $2.32\pm0.08$ (bb) &                        & \\
\hline
\end{tabular}
\begin{tablenotes}
    \item
\end{tablenotes}
\label{tab:alphaox}
\end{table}

In summary, we concluded that the 2021 flare is a TDE.


\section{The nature of the recurring flares}

Table~\ref{tab:two_flares} lists the basic properties of the two flares in F01004-2237.
The peak luminosity and energy budget of the 2010 flare were collected from \cite{Tadhunter2017} and \cite{Dou2017}, which were estimated using V-band data and MIR data, respectively.
Taking the intersection of the results from the two works, we obtained a peak luminosity of $4-11\times10^{44}$ erg s$^{-1}$ and a total energy of $1-11\times10^{52}$ erg, which are both higher than those of the 2021 flare.
The time interval of the peaks of the two flares is $10.3\pm0.3$ yr in the rest frame.
In this section, we discuss the possible origins of the flare's recurrence under the premise that the 2021 flare is TDE.

\begin{table}
\centering
\caption{The properties of the two flares.}
\begin{tabular}{ccc}
\hline
\hline
     & 2010 flare & 2021 flare  \\
\hline
$t_{\rm peak}$ (MJD)    & $55338\pm118$   & $59516\pm3$ \\
$m_{\rm peak}$          & $V=17.4\pm0.5$  & $c=17.42\pm0.02$ \\
\multirow{2}{*}{$L_{\rm peak}$ ($10^{44}$ erg s$^{-1}$)}  & 0.4--14 (T17) & \multirow{2}{*}{$4.4\pm0.4$}\\
                                                          & 4--11 (D17)   & \\
\multirow{2}{*}{$E_{\rm tot}$ ($10^{52}$ erg)}            & 0.3--11 (T17) & \multirow{2}{*}{$0.52\pm0.04$}\\
                                                          & $>$1--2 (D17) & \\
\hline
\end{tabular}
\begin{tablenotes}
    \item
\end{tablenotes}
\label{tab:two_flares}
\end{table}

\subsection{The nature of the 2010 flare} \label{sec:5.1}

As mentioned in the introduction section, \cite{Tadhunter2017} considered the 2010 flare as a TDE other than an AGN flare because of the He II BEL with large He II/H$\beta$ ratio in the spectrum $\sim$5 years after the flare occurred.
In recent years, a new class of BF AGN flares has been identified, and examples are AT 2017bgt, OGLE17aaj, AT 2021loi, and AT 2019aalc \citep{Trakhtenbrot2019,Gromadzki2019,Makrygianni2023,MilanVeres2024}.
These flares are characterized by a strong N III $\lambda$4640 BF emission line with similar intensity to He II $\lambda$4686, both commonly seen in TDEs.
\cite{Trakhtenbrot2019} and subsequent works demonstrated that these flares are not TDEs because their emission lines are narrower than typical TDEs and their LCs show double-peak structures, and suggested that they originate from a sudden enhancement of the long-existing accretion flow.
They also placed the 2010 flare in F01004-2237 into this class as they claimed similarity between the spectra of F01004-2237 and BF AGN flares.

However, the claimed similarity is questionable because there are two crucial differences that need to be explained\footnote{We caution readers that it may be misleading to compare the 2015 spectrum directly to those of BF flares without removing the long-standing WR features and NELs from the host galaxy as some works did.}.
One is that the BELs in F01004-2237 are broader and show a significant change in width.
The FWHMs of BELs are $\sim$6,000 km s$^{-1}$ in the 2015 spectrum \citep{Tadhunter2017}, and can reach 16,000--19,000 km s$^{-1}$ after the 2021 flare.
For comparison, the BEL widths of the BF flares are between 2,000 and 5,000 km s$^{-1}$ and do not change dramatically over time.
The other is that He II in F01004-2237 is stronger relative to H$\beta$.
In the 2015 spectrum, the He II/H$\beta$ ratio is $1.82\pm0.09$ if emission from all sources are included, and is $5.3\pm0.9$ if contributions from WR stars and ISM are removed \citep{Tadhunter2017}.
For comparison, the BF flares show He II/H$\beta$ ratios of 0.4--0.7.
The large and variable width and the large He II/H$\beta$ ratio of BELs are prevalent in TDEs, and BF lines are more common in TDEs than in AGNs \citep[e.g.][]{vanVelzen2020}.
Thus, we considered that the 2010 flare in F01004-2237 is more similar to TDEs than AGN flares.

We re-examined the nature of the 2010 flare in light of new observations of the 2021 flare.
If the 2010 flare were indeed an AGN flare as proposed by \cite{Trakhtenbrot2019}, the He II emission lines in the 2015 spectrum would indicate a broad line region with bulk velocities of several $10^3$ km s$^{-1}$.
The broad line region would still exist after the 2021 flare as the dynamic time scale is typically $>10$ yr, and would produce an emission component with a similar width as in 2015.
In fact, no such component in He II or other emission lines has been detected in early-time spectra taken after the 2021 flare.
The difference in the ionization continuum can hardly explain the no detection because the 5100 \AA\ luminosities at the time when these spectra were taken are $0.7-4\times10^{43}$ erg s$^{-1}$, similar to that in 2015.
Thus, the new observations do not support the AGN flare interpretation, and we preferred that a TDE causes the 2010 flare.
Note that the questions raised by \cite{Trakhtenbrot2019} about the TDE interpretation have been answered by \cite{Tadhunter2021} and \cite{Cannizzaro2021}.

\subsection{Repeating partial TDEs?} \label{sec:5.2}

If related, the recurring flares in F01004 may be caused by repeating partial TDEs or double TDEs.

After a partial TDE occurs, the stellar remnant may return, producing the next TDE.
The partial TDE of a star intruding in a parabolic orbit could transfer the stellar remnant into an elliptical orbit due to the weak asymmetry of the disrupted material.
However, the period would be $>400$ yr \citep{Ryu2020c}, which cannot explain the 10.3 yr interval between the two flares.
Therefore, repeating TDEs with an interval of 10.3 yr requires that the star was already in an elliptical orbit around the SMBH with a period of $\sim10$ years before the disruption.

The observed period corresponds to an orbital semi-major axis $a\sim1400$ AU for an SMBH mass of $\sim2.5\times10^7$ $M_\odot$.
For a partial TDE to occur, the pericenter $r_p$ must be similar to $r_t$, which is $\sim1.4$ AU for sun-like stars.
Then, the orbit must be highly eccentric with $1-e\sim10^{-3}$.
Such an orbit can be explained by the Hills mechanism.
If so, the orbital semi-major axis of the captured star is related to the initial orbital parameters of close binary stars \citep{Pfahl2005} as:
\begin{equation}
a \approx 0.35 a_b \left( \frac{m_b}{m_{\rm ej}} \right)  \left( \frac{M_{\rm BH}}{m_b} \right)^{2/3},
\end{equation}
where $a_b$ and $m_b$ are the binary semi-major axis and total mass, and $m_{\rm ej}$ is the mass of the ejected star.
Assuming both the binary stars have a mass of 1 $M_\odot$, the observed period corresponds to $a_b\approx8r_\odot$.
The inferred binary semi-major axis is reasonable as it is far enough to avoid a common envelope, and close enough to avoid disintegration in collisions with fellow stars \citep{Hills1988}.

Simulations show that the flare produced by a partial TDE has a steeper descent in the late-time, obeying $L\propto t^{-9/4}$ rather than $L\propto t^{-5/3}$ as in full TDEs \citep[e.g.,][]{Coughlin2019,Miles2020}.
We tested this with the observations of the 2021 flare in F01004-2237.
We fit the ATLAS/o, ATLAS/c and Swift UVW2 LCs after $+$60 day with t$^{-9/4}$ curves.
The procedure was similar to the fitting with t$^{-5/3}$ curves (Section~\ref{sec:3.2}), except that we replaced the powerlaw index in the equation (3) with $-9/4$.
This slightly improves the fit as the $\chi^2$/d.o.f. decreases from 3.57 to 3.31, with d.o.f.$=324$.
Therefore, t$^{-9/4}$ curves agree slightly better with the data than t$^{-5/3}$ curves, which supports the partial TDE model.

\subsection{Double TDEs of binary stars?} \label{sec:5.3}

There is a rich diversity of possible outcomes for the binary-SMBH encounter, with one possibility being that both stars are tidally disrupted \citep{Mandel2015,Mainetti2016}.
This phenomenon, known as double TDE, occurs when the pericenter of the binary's centre of mass is close enough to the black hole.
However, the time interval between the two TDEs is generally less than $\sim10^2$ days, resulting in only one flare being observed, which is inconsistent with the situation in F01004-2237.

In the double TDE regime, two possible scenarios could explain the interval of 10.3 yr.
One scenario was proposed by \cite{Mandel2015} to interpret the two flares 20 years apart in IC 3599.
A possible outcome of the binary-SMBH encounter is that one star is directly disrupted, causing the first TDE, and then the other star is captured into an elliptical orbit and produces the second TDE when it returns to the pericenter.
According to the simulations of \cite{Mandel2015}, the orbital period of the captured star can range from 6 months to several decades, explaining the observations in F01004-2237.

The other scenario involves the encounter between a stellar binary and a mpc-scale SMBHB binary.
As shown in the simulations by \cite{Wu2018} and \cite{Coughlin2018}, there is a significant probability of delayed double TDEs, among which the time interval of two TDEs can be more than ten years.
A mpc-scale secondary SMBH may pass through the stellar debris stream of the TDE as it orbits the primary SMBH and cause periodic gaps on the X-ray LC \citep{Liu2014,Shu2020}.
Unfortunately, the X-ray emission from the flare in F01004-2237 is weak and cannot be used to test the presence of mpc-scale secondary SMBH.

\subsection{Two independent flares?} \label{sec:5.4}

We considered the possibility that the two flares are not associated physically.
The probability of a TDE (the 2021 flare) occurring in a time interval $\Delta t$ after a specific event (the 2010 flare) is $p=r_{\rm TDE}\times\Delta t$, where $r_{\rm TDE}$ is the incidence rate of TDE per galaxy in a unit of yr$^{-1}$.
Adopting the rate of optical TDEs of $3.2^{+0.8}_{-0.6}\times10^{-5}$ yr$^{-1}$ \citep{Yao2023} and $\Delta t=10.3$ yr, we estimated a small probability of $p\sim3.3\times10^{-4}$.

For $p$ to increase to $>0.1$, the TDE rate needs to be $>10^{-2}$ yr$^{-1}$, $>300$ times higher than in normal galaxies.
Thus, assuming the two flares are independent requires an extremely high TDE rate in F01004-2237.
On the contrary, assuming that the two flares are repeating partial TDEs or double TDEs does not require such a high TDE rate because they should be treated as ``a single event'' when calculating the probability.

As there are many theories to explain the high TDE rate in post-starburst galaxies, we examined if they also apply to F01004-2237.
First, a starburst can cause an unusually high concentration of stars in the nucleus, enhancing the TDE rate \citep{Stone2016a,Stone2016b}.
In theory, this effect subsides over time \citep{Stone2018}.
The starburst in F01004-2237 started only 3--6 Myr ago, as estimated by the age of the WR stars \citep{Tadhunter2017}.
Based on the results of \cite{Stone2018}, we inferred that the TDE rate in F01004-2237 may be 1--2 orders of magnitude higher than that in post-starburst galaxies with stellar ages of $\sim$100 Myr, and hence 2--3 orders of magnitude higher than in normal galaxies.

Second, a nuclear star cluster (NSC) can enhance the TDE rate by orders of magnitude in galaxies with different masses and various types \citep{Pfister2020}.
The \textit{HST} images show a point-like source in the nucleus of F01004-2237, with SED consistent with a young stellar population with an age of $<10$ Myr and a mass of $10^8$ $M_\odot$ \citep{Surace1998}.
The source is barely resolved, indicating a $<100$ pc size.
This supports the existence of an NSC in F01004-2237, with mass and age unusual among NSCs in galaxies.
How much such an NSC can raise the TDE rate remains to be answered by theoretical studies.

Third, the presence of a companion SMBH with a mass ratio of 0.01--0.1 can significantly boost the TDE rate \citep[e.g.,][]{Ivanov2005}.
According to the simulations \citep[e.g.,][]{Chen2009,Li2017}, in a phase lasting for $\sim$Myr, during which the SMBHs are $\sim$pc apart, the TDE rate could be as high as $10^{-2}\sim1$ yr$^{-1}$.
Although there is no direct evidence, F01004-2237 may also host a pc-scale SMBH binary, given that starbursts are generally associated with mergers, which can lead to SMBH binaries.
There are no obvious merger signals in the \textit{HST} image \citep{Surace1998}.
However, this cannot negate the scenario of SMBH binary, because non-equal mass SMBH binaries may be related to minor mergers, and fall to the pc-scale long after mergers.

Therefore, a high TDE rate of $>10^{-2}$ yr$^{-1}$ in F01004-2237 is possible in theory.
As a result, we cannot rule out the possibility that the two flares are independent.

\section{Discussions}

\subsection{Evidence of pre-flare AGN claimed by literatures} \label{sec:6.1}

Some literature claimed to have found evidence for luminous AGN based on Seyfert-like optical narrow emission line (NELs) ratios \citep[e.g.,][]{Allen1991,Yuan2010}.
We noticed that the classification based on narrow emission lines is controversial in the literature, as \cite{Veilleux1999} classified it as star-forming.
The optical NELs show a narrow component with FWHM of $\sim$90 km s$^{-1}$ and broad components blueshifted by $\sim1000$ km s$^{-1}$, which is considered to be related with outflow \citep{Lipari2003,RodriguezZaurin2013,Tadhunter2021}.
\cite{RodriguezZaurin2013} shows that while the narrow component has starforming-like line ratios, the broadest component has Seyfert-like line ratios.
This may be why previous literature made contradictory classifications, as pointed out by \cite{Tadhunter2017}.
The outflow component that shows Seyfert-like line ratios has a radial extent to $\sim$3.5 kpc, as measured by \cite{RodriguezZaurin2013} using long-slit spectroscopy.
Meanwhile, it is not detected in the STIS 2000 spectrum extracted within an aperture corresponding to 350 pc in physical \citep{Farrah2005}.
Thus, its distance to the centre is $>350$ pc and thus traces AGN activity $\sim10^3-10^4$ years ago.
However, the levels of nuclear activity may vary dramatically on such a time scale, an example being Arp 187 \citep{Ichikawa2019}.
We are more concerned about recent activity, for which NELs in the nuclear spectrum are needed for diagnosis.
As measured by \cite{Farrah2005}, the NEL line ratios in the nuclear region fall around the classification lines between AGN and starforming \citep{Kewley2001,Kewley2006}, and classifications based on [S II]/H$\alpha$ and [O I]/H$\alpha$ are different.
Note that the result depends on the choice of classification lines; for example, the line ratios will be the explicit starforming type if the classification lines of \cite{Veilleux1987} are used.
The classification lines of Kewley et al. were calculated assuming the Salpeter's initial mass function, while the presence of a large population of WR stars in the nuclear region of F01004-2237 may result in line ratios closer to Seyfert galaxies than in normal star-forming galaxies.
Thus, spectral analysis in the nuclear region does not show evidence of strong AGN in recent $10^3$ years.
This is consistent with the absence of BELs and the weak hard X-ray emission.

The MIR spectrum of F01004-2237 shows only week PAH emission features with equivalent widths lower than expected for starburst galaxies \citep{Tran2001,Imanishi2007,Veilleux2009}.
This was considered to be evidence of AGN, whose intense X-ray and EUV emissions can destroy PAHs \citep[e.g.,][]{Voit1992}.
However, a weak AGN at present and a strong AGN in the past can also explain the destruction of PAH, and a strong AGN at present is not necessary.

F01004-2237 shows a hotter IR SED than starforming galaxies.
The IR SED indexes, such as flux ratios between 30 and 15 $\mu$m, and between MIR to FIR, are all consistent with ULIRGs with clear AGNs \citep{Veilleux2009}.
However, the MIR continuum of F01004-2237 shows only a weak 9.7 $\mu$m silicate absorption feature, and does not show strong dust temperature gradients \citep{Imanishi2007,Veilleux2009}, both of which are expected to be signatures of a buried AGN.
This leads us to suspect that the warm dust radiating the MIR continuum may have illumination sources other than the AGN.
As previously mentioned, F01004-2237 hosts an NSC with an age of $<10$ Myr, a mass of $10^8$ $M_\odot$, and a size of $<100$ pc.
Such an NSC may create a dense radiation field, producing a higher dust temperature than normal star-forming galaxies.
Moreover, considering that two energetic flares occurred in recent 14 years in F01004-2237, other flares possibly occurred in decades prior to the \textit{Spitzer} observations in 2004, although no observations were available to test them.
The 2010 flare was accompanied by a bright dust echo lasting for at least ten years, resulting in increases of monochromatic luminosities at 3.4 and 4.6 $\mu$m of $2-3\times10^{44}$ erg s$^{-1}$ \citep{Dou2017}.
If the hypothetic earlier flares also had similar total energies, they might significantly increase the temperature of the nuclear dust, producing an IR SED mimicking those of AGNs.
Therefore, it is not reliable to estimate the level of AGN with MIR luminosity.


\subsection{Comparison with known recurring flares} \label{sec:6.2}

\begin{table*}
\scriptsize
\centering
\caption{A summary of known energetic recurring flares.}
\begin{tabular}{ccccccccccc}
\hline
\hline
ID & z & log $M_{\rm BH}$ & $n_{\rm flare}$ & $\Delta t$ & $E_1$/$L_1$ & $E_2$/$L_2$ & $E_{\rm max}$ & $r_{21}$ & $t_{\rm next}$ & Ref. \\
(1)&(2)& (3)              & (4)             & (5)        & (6)         & (7)         & (8)           & (9)      & (10)           & (11) \\
\hline
ASASSN-14ko  & 0.0425 & 7.9 & $>$21 & 0.31  & \multicolumn{2}{c}{$L_{\rm UV}=1.8\times10^{44}$} & *$E_{\rm UV}\sim3\times10^{50}$ & $\sim1$ & - & 1,2,3 \\
\hline
eRASSt J0456 & 0.077  & 7.0 & 5 & $\sim$0.6 & $E_{\rm X}=1.0\times10^{52}$ & $E_{\rm X}=3.6\times10^{51}$ & $E_{\rm X}=1.0\times10^{52}$ & 0.4 & - & 4,5 \\
             &        &     &   &      & $L_{\rm X}=4.6\times10^{44}$ & $L_{\rm X}=1.1\times10^{44}$ & & & \\
\hline
AT 2022dbl   & 0.0284 & 6.4 & 2 & 1.89 & $L_{\rm UV}=7.8\times10^{43}$& $L_{\rm UV}=3.0\times10^{43}$& *$E_{\rm UV}\sim3\times10^{50}$ & 0.4 & 2026/01 & 6 \\
\hline
AT 2020vdq   & 0.045  & 6.1 & 2 & 2.54 & $E_{\rm UV}=6\times10^{49}$  & $E_{\rm UV}=2\times10^{51}$  & $E_{\rm UV}=2\times10^{51}$ & 30 & 2026/01 & 7 \\
             &        &     &   &      & $L_{\rm UV}=6\times10^{42}$  & $E_{\rm UV}=1\times10^{44}$  & & & \\
\hline
AT 2018fyk   & 0.059  & 7.7 & 2 & 3.1  & $L_{\rm UV}=3\times10^{44}$  & $L_{\rm UV}=7\times10^{42}$  & $E_{\rm UV}=9\times10^{51}$ & 0.02 & 2025/03 & 8,9 \\
\hline
F01004-2237  & 0.1178 & 7.4 & 2 & 10.3 & $E_{\rm UV}=\sim1-11\times10^{52}$ & $E_{\rm UV}=0.5\times10^{52}$ & $E_{\rm UV}>1\times10^{52}$ & 0.05--0.5 & 2033 & 10,11 \\
             &        &     &   &      & $L_{\rm UV}=\sim4-11\times10^{44}$ & $L_{\rm UV}=4.4\times10^{44}$ & & & \\
\hline
IC 3599      & 0.0215 & 6.7 & 2 & 18.8 & $L_{\rm X}>5.6\times10^{43}$ & $L_{\rm X}>1.5\times10^{43}$ & *$E_{\rm X}\gtrsim4\times10^{50}$ & 0.3? & 2029 & 12,13,14 \\
\hline
RX J1331-3243& 0.0519 & 6.5 & 2 & 27.5 & $L_{\rm X}>1\times10^{43}$   & $L_{\rm X}>6\times10^{42}$   & - & 0.6? & 2050 & 15,16 \\
\hline
\end{tabular}
\begin{tablenotes}
    \item (1): The identification of the flares (eRASSt J045650.3-203750 and RX J133157.6-324319.7 are listed as short names).
    (2): Redshift. (3): Black hole mass in a unit of $M_\odot$. For IC 3599, we adopted value from \cite{Grupe2015}.
    (4): Number of flares reported. Note that we did not adopt the three-flares model for IC 3599 proposed by \cite{Campana2015} because the predicted flare in 2019 was not seen \citep{Grupe2024}.
    (5): The time interval of the observed peaks of the flares in a unit of years in the rest frame.
    (6)(7): The total energies (if present in the literature) and the peak luminosities of the first and the second flares. We list the energies and luminosities in 0.2--2 keV for those flares detected in the X-ray band, and the values in the UV band from blackbody fit for other flares. For ASASSN-14ko, all flares have similar luminosities, so we list the average for the flares studied in detail. For IC 3599 and RX J1331-3243 with sparse data sampling, we adopted the observed peak luminosities as lower limits.
    (8): The total energy of the most energetic flare. The values with a ``*'' label are our own estimates based on literature data, which is not accurate and only used for order of magnitude estimates. Others are taken directly from the literature. The total energy of flares in RX J1331-3243 cannot be estimated due to a lack of observational data.
    (9): The ratio between the energies of the second and the first flares. If energies are absent in the literature, we calculated the ratio between peak luminosities. Note that the ratios are only for reference for IC 3599 and RX J1331-3243 with only lower limits of luminosities.
    (10): The predicted peak time of the next flare for those occurring only twice.
    (11): References are: 1. \cite{Payne2021}; 2. \cite{Payne2022}; 3. \cite{Payne2023}; 4. \cite{Liu2023}; 5. \cite{Liu2024}; 6. \cite{Lin2024}; 7. \cite{Somalwar2023}; 8. \cite{Wevers2019}; 9. \cite{Wevers2023}; 10. \cite{Tadhunter2017}; 11. \cite{Dou2017}; 12. \cite{Grupe1995}; 13. \cite{Grupe2015}; 14. \cite{Campana2015}; 15. \cite{Hampel2022}; 16. \cite{Malyali2023}.
\end{tablenotes}
\label{tab:recurring}
\end{table*}

In this section, we compared the recurring flares in F01004-2237 and other galaxies.
We check whether all of them can be interpreted under the framework of repeating partial TDEs, as this is how most of the literature explains them.
The information on these flares is listed in Table~\ref{tab:recurring}.
X-ray QPEs are not listed here because their time scales, luminosities and energy budgets are quite different from flares in F01004-2237.

In the scenario of repeating partial TDEs formed by the Hills mechanism, the outcome of the captured star depends on parameter $\beta \equiv r_t/r_p$.
The star is partially disrupted if $\beta$ is greater than the critical value and is not disrupted otherwise.
Depending on $\beta$, repeating partial TDE can be divided into two groups.
In one group, $\beta$ is only slightly above the critical value, causing only a tiny fraction of stellar mass to be disrupted, leaving flares with low energies and short durations \citep[e.g.,][]{Ryu2020c,Nixon2021}.
Since a partial TDE has little effect on $r_p$ and the stellar mass decreases only a little, $\beta$ is almost constant and similar partial TDEs can repeat many times.
In the other group, $\beta$ is significantly larger than the critical value, causing a large fraction to be disrupted, resulting in energetic flares.
Only several flares may be detected because the remaining flares are too weak due to small residual stellar mass.

This scenario can explain the diversity of the observational features of recurring flares.
We divided all recurring flares into two groups based on the above analysis.
The first group include ASASSN-14ko and X-ray QPEs, where $>9$ flares were detected, and a single flare has a UV energy $<10^{51}$ erg or an X-ray energy $<10^{50}$ erg.
The second group include the other recurring flares.
We preferred that the difference in $\beta$ is decisive, while the other differences are not essential.
The two groups appear to have different periods $P$ ($P<114$ days for the first group and $P>223$ days for the second group).
However, this might be a selection effect.
Long-period, low-energy repeating partial TDEs might exist but be extremely difficult to identify.
Short-period, energetic partial TDEs might also exist but be misclassified as a single TDE because the flares overlap.
In addition, most flares in the first group were detected in the X-ray band, while flares in the second group could be detected in both the UV/optical and X-ray bands.
The possible reason is that a larger debris mass leads to a higher chance of forming an optically thick envelop through debris collision or outflow, producing strong UV/optical emission \citep[e.g.,][]{Dai2018}.

In the scenario of repeating partial TDEs, will the debris mass of the next TDE be greater than, less than, or roughly equal to that of the previous one?
It is challenging to give an exact answer because the remnant fraction has a complex relationship with the stellar mass and $\beta$ parameters, as shown in Fig. 4 of \cite{Ryu2020c}.
To complicate matters further, stellar remnants that survive a partial TDE do not have enough time to reach equilibrium before returning \citep{Ryu2020c}, and hence, when calculating the remnant fraction of the next TDE, the results of stars of the same mass cannot be simply reproduced.
What is clear, however, is that the debris in the next TDE is unlikely to be much more massive than in the previous one.
This is because if the stellar core survives a partial TDE as only (part of) the stellar envelope lost, its structure changes little before returning to the pericenter since the time scale of thermal relaxation is much longer than the orbital period \citep{Ryu2020c}, and thus it is likely to survive the next partial TDE.

We tested this with observational data.
Most recurring flares meet theoretical expectations as the next flare is weaker than or similar to the previous one in total energy and peak luminosity.
The only exception is AT 2020vdq, where the second flare is 15 times brighter than the first and has total energy 30 times higher \citep{Somalwar2023}.
This is difficult to understand in the scenario of repeating partial TDEs.
Meanwhile, in the double TDEs scenario, two flares are produced by two different stars in the binary system, and there is no direct correlation between their luminosities and energies.
In addition, only two flares have been detected so far for AT 2020vdq, not contradicting a double TDEs scenario.
Thus, AT 2020vdq may be a case of double TDEs, rather than repeating partial TDEs.

If AT 2020vdq is double TDEs, then the other recurring flares appearing only twice may also be double TDEs, because situations where the second flare is stronger than, weaker than or similar to the first one can all occur in the double TDEs scenario.
However, it is more likely that ASASSN-14ko and eRASSt J045650.3-203750 are repeating partial TDEs because they show multiple flares.
Therefore, we proposed that the recurring flares detected so far could not be explained with only one scenario, and repeating partial TDEs and double TDEs may both exist.

\subsection{Future predictions} \label{sec:6.3}

For sources that have flared only twice currently, the best way to distinguish between the repeating partial TDEs and the double TDEs scenarios is to wait and check if there is an expected third flare.
We list the timing of these expected flares in the repeating partial TDEs scenario in Table~\ref{tab:recurring}.
The next flare expected in F01004-2237 will peak between December 2032 and August 2033.
However, it may come earlier, considering that the flare interval in eRASSt J045650.3-203750 is gradually shortening \citep{Liu2024}.
A flare that arrives at the scheduled time and is weaker than the previous one supports an interpretation of repeating partial TDEs, while the nondetection of the expected flare supports interpretations of double TDEs or independent TDEs.
Before that, AT 2018fyk, AT 2020vdq and AT 2022dbl will go to trials in 2025 and 2026.

In section~\ref{sec:5.4}, we demonstrated that a high TDE rate of $>10^{-2}$ in F01004-2237 is required if the two flares are independent.
Theoretically, a galaxy with a young starburst and massive NSC, like F01004-2237, may have a high TDE rate.
This can be tested by future monitoring of a large number of WR galaxies \citep[e.g.,][]{Brinchmann2008} with similar properties.
With the advent of new generation sky surveys such as the 2.5-meter Wide Field Survey Telescope \citep{Wang2023} and the Large Synoptic Survey Telescope \citep{LSST_paper}, TDEs with magnitudes $<23$ can be detected, meaning that the monitoring we mentioned earlier can be performed in $z<1$.
Within this large volume, there are at least a few hundred galaxies with properties similar to F01004-2237, taking into account the cosmological evolution of the star formation rate.
Monitoring them over several years is sufficient to prove or disprove the possible high TDE rate.

\section{Summary and conclusions}

A decade after the flare in 2010, a second optical flare was discovered on September 4, 2021, by ATLAS with $c=20.2$ in F01004-2237.
The flare was later independently detected by \textit{Gaia}, which showed that the flare's position coincides with the galaxy centre.
The flare peaks in $\sim50$ days with $o=17.64\pm0.02$, $c=17.42\pm0.02$, and $g=17.36\pm0.07$, corresponding to absolute magnitudes of $\sim-21$.
After the peak, the flare drops roughly as $L\propto t^{-5/3}$ and falls below the detection limit in all bands after $+$700 day.
Between $+$99 and $+$219 day, the flare maintains a nearly constant blackbody temperature of $\sim$22,000 K despite declining luminosity.
All these features are consistent with TDEs and inconsistent with SNe.
The multi-band LCs can be well fit by a TDE model, yielding a peak luminosity of $4.4\pm0.4\times10^{44}$ erg s$^{-1}$, and a total energy of $5.2\pm0.4\times10^{51}$ erg released in 700 days.

The optical and UV spectra of the flare taken between $+$95 and $+$325 day show VBELs with FWHM $\gtrsim6,000$ km s$^{-1}$, detected in H$\alpha$, H$\beta$, He II $\lambda$4686, He I $\lambda$5876 and Ly$\alpha$.
H$\alpha$ and He II VBELs have been continuously observed over more than 200 days, with H$\alpha$ FWHM varying from 7,000 to 19,000 km s$^{-1}$ and He II FWHM reaching $\gtrsim21,000$ km s$^{-1}$, and He II/H$\alpha$ ratios ranging from 0.3 to 2.3.
Such VBELs must come from the vicinity of the SMBH, and a high Ly$\alpha$/H$\alpha$ ratio of $10.4\pm0.4$ indicates little dust reddening in the line of sight.
The large FWHMs and the high He II/H$\alpha$ ratios of the VBELs strongly support a TDE nature and can hardly be explained with SNe and AGN flares.

Most of the time after the 2021 flare, no X-ray emission clearly associated with the flare was detected.
An \textit{XMM-Newton} observation on $+$247 day shows that the X-ray emission is weak relative to the UV emission as $\alpha_{\rm OX}>1.9$, a value much higher than $1.25\pm0.15$ in AGNs with similar UV luminosity.
Around $+$280 and $+$350 day, two X-ray flares were detected by \textit{Swift}/XRT with confidences of 3.5$\sigma$ and 2.5$\sigma$, respectively.
The flares last no more than 2--3 weeks, during which the X-ray spectrum is soft and can be described by a power-law with $\Gamma=4.35^{+1.44}_{-1.29}$ or a blackbody with $kT=113^{+40}_{-28}$ eV with no additional absorption beyond the milky way.
The soft, highly variable X-ray emission, which is weak relative to UV emission, supports a TDE nature.

Based on these observations, we conclude that a TDE caused the 2021 flare.
In addition, as no BEL components with several $10^3$ km s$^{-1}$ are seen in the new spectra, we prefer that the 2010 flare is also caused by a TDE.
The time interval between the two flares is $10.3\pm0.3$ yr, and the energy released in the second flare is 0.05--0.5 times that in the first.

If physically related, the recurring flares in F01004-2237 could be explained by either repeating partial TDEs or double TDEs.
We cannot distinguish between the two possibilities solely based on the observations of the two flares because theoretical works have yet to provide a straightforward means of distinguishing between them, at least until now.
The repeating partial TDEs scenario predicts that the next flare will arrive in 2032 or 2033.
Whether the flare occurs as scheduled can test the scenario.

If not physically related, the recurring flares require an extremely high TDE rate of $\gtrsim10^{-2}$.
Such a rate is theoretically possible given that F01004-2237 has just experienced a starburst and that there is a young star cluster with an age of 3--6 Myr and a mass of $10^8$ $M_\odot$, and could be tested by future photometric monitoring of galaxies with similar properties at $z<1$.

\begin{acknowledgements}

We acknowledge the anonymous referee for helping us improve this manuscript.
We thank for useful suggestions from Fukun Liu and Takayuki Hayashi.
We thank Minghao Yue for performing the Magellan observation.
This work is supported by the National Natural Science Foundation of China (NFSC, 12103002).
N.J. acknowledges the support of the NFSC (12073025, 12192221), the National Key R\&D Program of China (2023YFA1608100), the Strategic Priority Research Program of the Chinese Academy of Sciences (XDB0550200).
We thank the Swift science operations team for accepting our ToO requests and arranging the observations.
This work made use of data supplied by the UK Swift Science Data Centre (UKSSDC) at the University of Leicester.
Some of the observations reported in this paper were obtained with the Southern African Large Telescope (SALT), under program 2021-2-LSP-001 (PI: Buckley).
Polish participation in SALT is funded by grant No. MEiN nr 2021/WK/01.
Based on observations obtained at the international Gemini Observatory, a program of NSF NOIRLab, which is managed by the Association of Universities for Research in Astronomy (AURA) under a cooperative agreement with the U.S. National Science Foundation on behalf of the Gemini Observatory partnership: the U.S. National Science Foundation (United States), National Research Council (Canada), Agencia Nacional de Investigaci\'{o}n y Desarrollo (Chile), Ministerio de Ciencia, Tecnolog\'{i}a e Innovaci\'{o}n (Argentina), Minist\'{e}rio da Ci\^{e}ncia, Tecnologia, Inova\c{c}\~{o}es e Comunica\c{c}\~{o}es (Brazil), and Korea Astronomy and Space Science Institute (Republic of Korea).
This research uses data obtained through the Telescope Access Program (TAP), which has been funded by the TAP member institutes.
Observations obtained with the Hale Telescope at Palomar Observatory were obtained as part of an agreement between the National Astronomical Observatories, Chinese Academy of
Sciences, and the California Institute of Technology.

\end{acknowledgements}

\bibliographystyle{aa}
\bibliography{F01004_2021flare}

\begin{appendix}

\section{A list of the \textit{Swift} observations and data}

\setcounter{table}{0}
\renewcommand{\thetable}{A\arabic{table}}

\begin{table*}
\centering
\caption{\textit{Swift} observation log and XRT count rates}
\begin{tabular}{ccccccccc}
\hline
\hline
OBSID & date & MJD & phase(day) & $t_{\rm exp}$ (s) & $N_{\rm tot}$ & $N_{\rm bkg}$ & $f_c$ & rate ($10^{-3}$ cnt s$^{-1}$)\\
\hline
00014990001 & 2021-12-24 & 59572.4 &  99.1 & 1569.6 &  4 &  1.584 & 1.352 & $< 6.68$\\
00014990002 & 2021-12-30 & 59578.5 & 104.6 & 1654.8 &  5 &  0.297 & 1.467 & $ 4.17^{+2.33}_{-1.72}$\\
00014990003 & 2022-01-07 & 59586.6 & 111.8 & 1557.0 &  4 &  0.341 & 1.897 & $ 4.46^{+2.93}_{-2.08}$\\
00014990004 & 2022-01-15 & 59594.1 & 118.6 & 2166.3 &  3 &  0.329 & 2.391 & $ 2.95^{+2.37}_{-1.60}$\\
00014990008 & 2022-05-06 & 59705.9 & 218.6 & 6376.1 &  7 &  0.663 & 1.467 & $ 1.46^{+0.70}_{-0.54}$\\
00014990009 & 2022-05-21 & 59720.8 & 232.0 & 2008.3 &  2 &  0.218 & 2.274 & $ 2.02^{+2.10}_{-1.28}$\\
00014990010 & 2022-06-04 & 59734.7 & 244.3 & 1499.4 &  1 &  0.264 & 1.380 & $< 4.26$\\
00014990011 & 2022-06-17 & 59747.7 & 256.0 & 1960.7 &  1 &  0.169 & 1.560 & $< 3.74$\\
00014990012 & 2022-07-02 & 59762.8 & 269.5 &  860.0 &  2 &  0.302 & 1.374 & $< 9.79$\\
00014990013 & 2022-07-16 & 59776.3 & 281.5 & 1033.0 & 13 &  0.375 & 1.328 & $16.23^{+5.11}_{-4.24}$\\
00014990014 & 2022-07-30 & 59790.5 & 294.3 & 1511.9 &  6 &  0.369 & 1.380 & $ 5.14^{+2.59}_{-1.97}$\\
00014990015 & 2022-08-13 & 59804.1 & 306.4 & 1599.7 &  1 &  0.177 & 4.308 & $<12.65$\\
00014990016 & 2022-08-27 & 59818.7 & 319.5 & 1589.6 &  2 &  0.233 & 1.765 & $ 1.96^{+2.06}_{-1.26}$\\
00014990017 & 2022-09-10 & 59832.6 & 331.9 & 1384.0 &  6 &  0.593 & 1.422 & $ 5.56^{+2.92}_{-2.21}$\\
00014990018 & 2022-09-24 & 59846.5 & 344.4 &  990.4 &  8 &  0.313 & 1.355 & $10.52^{+4.39}_{-3.46}$\\
00014990019 & 2022-10-02 & 59854.6 & 351.6 & 1639.8 & 11 &  0.541 & 1.347 & $ 8.59^{+3.03}_{-2.48}$\\
00014990020 & 2022-10-05 & 59857.8 & 354.5 & 1702.5 &  2 &  0.199 & 1.902 & $ 2.01^{+2.07}_{-1.27}$\\
00014990021 & 2022-10-08 & 59860.3 & 356.7 & 1459.2 &  6 &  0.397 & 1.547 & $ 5.94^{+3.01}_{-2.28}$\\
00014990023 & 2022-10-14 & 59867.0 & 362.7 & 1206.0 &  2 &  0.125 & 1.572 & $ 2.44^{+2.42}_{-1.48}$\\
00014990024 & 2022-10-22 & 59874.4 & 369.4 &  451.3 &  1 &  0.061 & 1.477 & $<15.71$\\
00014990025 & 2022-10-27 & 59879.9 & 374.2 & 1830.3 &  5 &  0.301 & 1.582 & $ 4.06^{+2.28}_{-1.68}$\\
00014990026 & 2022-10-30 & 59882.5 & 376.6 & 1532.0 &  2 &  0.175 & 1.506 & $ 1.79^{+1.82}_{-1.12}$\\
00014990027 & 2022-11-03 & 59886.6 & 380.2 & 1557.0 &  5 &  0.283 & 1.614 & $ 4.89^{+2.73}_{-2.01}$\\
00014990028 & 2022-11-06 & 59889.9 & 383.2 & 2061.0 &  2 &  0.253 & 1.450 & $ 1.23^{+1.30}_{-0.80}$\\
00014990030 & 2022-11-15 & 59898.6 & 390.9 &  742.2 &  3 &  0.143 & 1.429 & $ 5.50^{+4.14}_{-2.79}$\\
00014990031 & 2022-11-18 & 59902.0 & 394.0 & 1080.6 &  3 &  0.206 & 1.738 & $ 4.49^{+3.46}_{-2.33}$\\
00014990032 & 2022-11-23 & 59906.2 & 397.8 & 2111.1 &  4 &  0.279 & 2.216 & $ 3.91^{+2.52}_{-1.80}$\\
00014990033 & 2023-05-07 & 60071.4 & 545.5 & 3089.0 &  3 &  0.442 & 2.547 & $ 2.11^{+1.77}_{-1.19}$\\
00097667001 & 2024-05-06 & 60436.9 & 872.5 & 1973.2 &  2 &  0.226 & 1.499 & $ 1.35^{+1.41}_{-0.86}$\\
\hline
\end{tabular}
\label{tab:xrtdata}
\end{table*}

\begin{table*}
\small
\centering
\caption{Photometries of \textit{Swift}/UVOT and \textit{XMM-Newton}/OM}
\begin{tabular}{cccccccc}
\hline
\hline
OBSID & MJD & UVW2 & UVM2 & UVW1 & U & B & V\\
\hline
\multicolumn{8}{c}{\textit{Swift}/UVOT}\\
\hline
00014990001 & 59572.4 & $17.149\pm0.046$ & $17.126\pm0.046$ & $17.192\pm0.052$ & $17.219\pm0.064$ & $17.270\pm0.083$ & $16.973\pm0.115$ \\
00014990002 & 59578.5 & $17.140\pm0.046$ & $17.095\pm0.046$ & $17.179\pm0.052$ & $17.244\pm0.063$ & $17.336\pm0.083$ & $17.098\pm0.126$ \\
00014990003 & 59586.6 & $17.245\pm0.048$ & $17.195\pm0.046$ & $17.244\pm0.055$ & $17.184\pm0.066$ & $17.303\pm0.089$ & $17.177\pm0.145$ \\
00014990004 & 59594.1 & $17.274\pm0.045$ & $17.206\pm0.045$ & $17.310\pm0.050$ & $17.269\pm0.059$ & $17.319\pm0.077$ & $17.010\pm0.108$ \\
00014990008 & 59705.9 & $17.430\pm0.041$ & $17.526\pm0.041$ & $17.624\pm0.044$ & $17.513\pm0.050$ & $17.574\pm0.070$ & $17.350\pm0.098$ \\
00014990009 & 59720.8 &                  &                  &                  & $17.570\pm0.033$ &                  &                  \\
00014990010 & 59734.7 & $17.540\pm0.048$ & $17.655\pm0.054$ & $17.629\pm0.050$ &                  &                  &                  \\
00014990011 & 59747.7 & $17.596\pm0.046$ & $17.627\pm0.050$ & $17.776\pm0.049$ &                  &                  &                  \\
00014990012 & 59762.8 & $17.730\pm0.057$ & $17.877\pm0.068$ & $17.809\pm0.062$ &                  &                  &                  \\
00014990013 & 59776.3 & $17.822\pm0.055$ & $18.054\pm0.068$ & $18.011\pm0.067$ &                  &                  &                  \\
00014990014 & 59790.5 & $17.909\pm0.050$ & $17.947\pm0.058$ & $17.984\pm0.056$ &                  &                  &                  \\
00014990015 & 59804.1 & $18.025\pm0.052$ & $18.103\pm0.060$ & $18.129\pm0.056$ &                  &                  &                  \\
00014990016 & 59818.7 & $17.912\pm0.042$ &                  &                  &                  &                  &                  \\
00014990017 & 59832.6 & $17.944\pm0.043$ &                  &                  &                  &                  &                  \\
00014990018 & 59846.5 & $18.033\pm0.046$ &                  &                  &                  &                  &                  \\
00014990019 & 59854.6 & $18.009\pm0.042$ &                  &                  &                  &                  &                  \\
00014990020 & 59857.8 & $17.997\pm0.042$ &                  &                  &                  &                  &                  \\
00014990021 & 59860.3 & $18.156\pm0.044$ &                  &                  &                  &                  &                  \\
00014990023 & 59867.0 & $18.044\pm0.044$ &                  &                  &                  &                  &                  \\
00014990024 & 59874.4 & $18.098\pm0.055$ &                  &                  &                  &                  &                  \\
00014990025 & 59879.9 & $18.039\pm0.042$ &                  &                  &                  &                  &                  \\
00014990026 & 59882.5 & $18.018\pm0.043$ &                  &                  &                  &                  &                  \\
00014990027 & 59886.6 & $18.039\pm0.043$ &                  &                  &                  &                  &                  \\
00014990028 & 59889.9 & $18.002\pm0.041$ &                  &                  &                  &                  &                  \\
00014990030 & 59898.6 & $18.012\pm0.057$ &                  &                  &                  &                  &                  \\
00014990032 & 59906.2 & $18.060\pm0.041$ &                  &                  &                  &                  &                  \\
00014990033 & 60071.4 & $18.271\pm0.041$ &                  &                  &                  &                  &                  \\
00097667001 & 60436.9 & $18.325\pm0.053$ & $18.272\pm0.070$ & $18.352\pm0.070$ & $18.327\pm0.088$ & $18.128\pm0.128$ & $17.721\pm0.172$ \\
\hline
\multicolumn{8}{c}{\textit{XMM-Newton}/OM}\\
\hline
0605880101 & 55173.7 & $18.439\pm0.087$ & $18.274\pm0.026$ & $18.373\pm0.017$ & $18.492\pm0.026$ & $18.376\pm0.028$ &                  \\
0911791101 & 59738.1 &                  & $17.621\pm0.018$ & $17.696\pm0.009$ & $17.814\pm0.009$ &                  &                  \\
\hline
\end{tabular}
\begin{tablenotes}
   \item (a) All the photometries are in the AB system and have been corrected for extinction from the Milky Way.
   \item (b) There is something wrong with UVOT data in observation 00014990031.
\end{tablenotes}
\label{tab:uvotdata}
\end{table*}

\section{Modelling of the pre-flare optical spectrum} \label{App:optspec}

\setcounter{figure}{0}
\renewcommand{\thefigure}{B\arabic{figure}}

\begin{figure*}
\centering
  \includegraphics[scale=0.9]{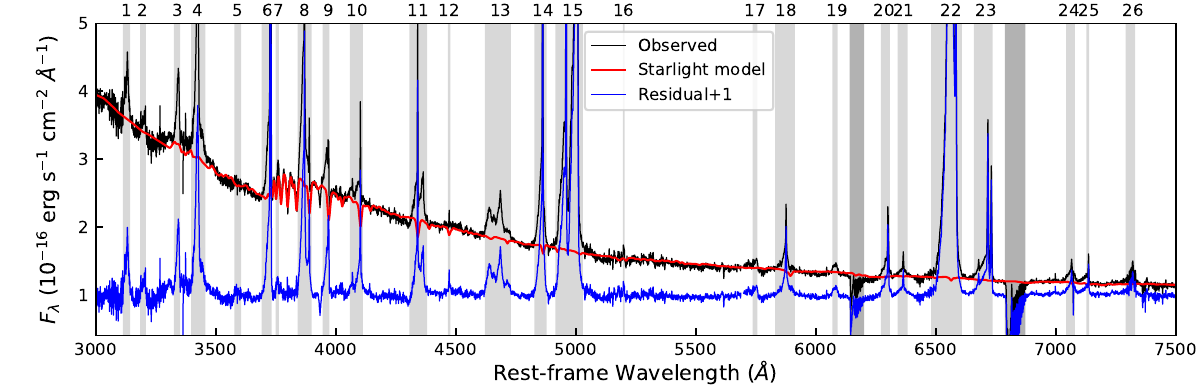}
  \caption{
  The best-fitting stellar model in the pre-flare spectrum.
  The grey shades label the emission lines identified by us.
  These are: 1. O III $\lambda\lambda$3121,3132; 2. He I $\lambda$3189; 3. [Ne V] $\lambda$3346; 4. [Ne V] $\lambda$3426; 5. [Fe VII] $\lambda$3586; 6. [O II] $\lambda\lambda$3726,3729; 7. [Fe VII] $\lambda$3759; 8. [Ne III] $\lambda$3869 + He I $\lambda$3889; 9. [Ne III] $\lambda$3968 + H$\epsilon$; 10. [S II] $\lambda$4076 + H$\delta$; 11. H$\gamma$ + [O III] $\lambda$4363; 12. He I $\lambda$4471; 13. N III $\lambda$4640 + He II $\lambda$4686; 14. H$\beta$; 15. [O III] $\lambda\lambda$4959,5007; 16. [N I] $\lambda$5199; 17. [Fe VII] $\lambda$5722 + [Fe II]$\lambda$5750??; 18. He I $\lambda$5876; 19. [Fe VII] $\lambda$6087; 20. [O I] $\lambda$6300; 21. [O I] $\lambda$6364 + [Fe X] $\lambda$6376; 22. H$\alpha$ + [N II] $\lambda$6548,6583; 23. He I $\lambda$6679 + [S II] $\lambda\lambda$6716,6731; 24. He I $\lambda$7065; 25. [Ar III] $\lambda$7136; 26. [O II] $\lambda\lambda$7320,7330.
  The dark grey shades label the regions affected by strong telluric absorption lines, which were not included in the fitting.
  }
\label{fig:starlight}
\end{figure*}

To interpret the flare's properties using the post-flare spectra, we must first model the host galaxy's contribution using the pre-flare spectrum.
We used the pre-flare spectrum taken by VLT/XShooter in August 2018, when the 2010 flare had faded.
The pre-flare spectrum shows a stellar continuum and numerous emission lines, which come from the host galaxy and the light echo of the 2010 flare \citep{Tadhunter2021}.
We only modelled the stellar continuum and did not model the emission lines because the latter can vary significantly due to varying observational conditions.
Instead, we masked the wavelength ranges affected by the emission lines.
The stellar model is a linear combination of the first eight principal components of the simple stellar population spectra provided by \cite{BC03}, which is broadened by a Gaussian function to match the stellar velocity dispersion and reddened following the extinction law of \cite{Calzetti1994} which applies for starburst galaxies.
To identify emission lines, we employed an iterative algorithm: we labelled the wavelengths with large residuals as emission lines and did not include these ranges in the fitting in the next iteration.
The final stellar model is shown in Fig.~\ref{fig:starlight}, where the identified emission lines are marked with grey shades.

\section{VBEL parameters}

\setcounter{table}{0}
\renewcommand{\thetable}{C\arabic{table}}

\begin{table}
\centering
\caption{Parameters of the VBELs}
\begin{tabular}{cccc}
\hline
\hline
  & Shifted velocity & FWHM        & Luminosity \\
  & km s$^{-1}$    & km s$^{-1}$ & $10^{41}$ erg s$^{-1}$ \\
\hline
\multicolumn{4}{c}{Magellan $+$108 day (Poly)}\\
\hline
H$\alpha$  &$ -551\pm 148$ & $19238\pm 438$ & $12.09\pm0.26$ \\
H$\beta$   &$ -233\pm1541$ & $18102\pm1704$ & $ 3.45\pm1.08$ \\
He II      &$ 7248\pm2059$ & $25126\pm3824$ & $ 4.17\pm1.11$ \\
He I       &$  835\pm 506$ & $15038\pm1027$ & $ 2.04\pm0.18$ \\
\hline
\multicolumn{4}{c}{Magellan $+$108 day (DBB)}\\
\hline
H$\alpha$  &$ -402\pm 150$ & $16215\pm 391$ & $ 9.52\pm0.22$ \\
H$\beta$   &$ 2337\pm 302$ & $17697\pm1272$ & $ 3.26\pm0.33$ \\
He II      &$ 3088\pm 998$ & $>23080$       & $ 7.02\pm0.47$ \\
He I       &$  494\pm 930$ & $10443\pm1942$ & $ 0.64\pm0.15$ \\
\hline
\multicolumn{4}{c}{Magellan $+$108 day (DPL)}\\
\hline
H$\alpha$  &$ -343\pm 155$ & $16184\pm 395$ & $ 9.50\pm0.23$ \\
H$\beta$   &$ 2367\pm 353$ & $16775\pm1405$ & $ 3.29\pm0.26$ \\
He II      &$ 2032\pm 954$ & $>22802$       & $ 6.93\pm0.42$ \\
He I       &$ 1020\pm1120$ & $ 8765\pm2498$ & $ 0.46\pm0.15$ \\
\hline
\multicolumn{4}{c}{Gemini-S $+$271 day (Poly)}\\
\hline
H$\alpha$  &$ -412\pm 229$ & $18111\pm 572$ & $ 4.30\pm0.14$ \\
H$\beta$   & -             & -              & $<1.31$ \\
He II      &$ -707\pm 943$ & $11074\pm2162$ & $ 1.40\pm0.32$ \\
\hline
\multicolumn{4}{c}{Gemini-S $+$271 day (DBB)}\\
\hline
H$\alpha$  &$  280\pm 248$ & $18501\pm 648$ & $ 4.34\pm0.12$ \\
H$\beta$   & -             & -              & $<1.63$ \\
He II      &$-2395\pm 910$ & $14826\pm2925$ & $ 2.52\pm0.45$ \\
\hline
\multicolumn{4}{c}{Gemini-S $+$271 day (DPL)}\\
\hline
H$\alpha$  &$  262\pm 246$ & $18634\pm 589$ & $ 4.37\pm0.12$ \\
H$\beta$   & -             & -              & $<1.63$ \\
He II      &$-2318\pm 831$ & $14476\pm2200$ & $ 2.53\pm0.37$ \\
\hline
\multicolumn{4}{c}{P200 $+$325 day (Poly)}\\
\hline
H$\alpha$  &$ -835\pm 352$ & $11140\pm1129$ & $ 1.41\pm0.11$ \\
H$\beta$   & -             & -              & $<0.59$ \\
He II      &$-3201\pm1771$ & $21881\pm3658$ & $ 1.36\pm0.22$ \\
\hline
\multicolumn{4}{c}{P200 $+$325 day (DBB)}\\
\hline
H$\alpha$  &$-1122\pm 286$ & $ 8255\pm1563$ & $ 1.29\pm0.16$ \\
H$\beta$   & -             & -              & $<0.51$ \\
He II      &$ -822\pm1817$ & $27933\pm3598$ & $ 1.98\pm0.29$ \\
\hline
\multicolumn{4}{c}{P200 $+$325 day (DPL)}\\
\hline
H$\alpha$  &$-1272\pm 219$ & $ 6724\pm 925$ & $ 1.24\pm0.20$ \\
H$\beta$   & -             & -              & $<0.49$ \\
He II      &$ 1696\pm1187$ & $38278\pm3626$ & $ 2.89\pm0.26$ \\
\hline
\end{tabular}
\begin{tablenotes}
   \item The shifted velocity is negative for blueshifted and positive for redshifted
   \item For none detections, we list the upper limit of the luminosity at 99.73\% probability.
\end{tablenotes}
\label{tab:par_vbel}
\end{table}

\section{Identification of UV absorption lines} \label{App:UVabs}

\setcounter{figure}{0}
\renewcommand{\thefigure}{D\arabic{figure}}

\begin{figure}
\centering
  \includegraphics[scale=0.7]{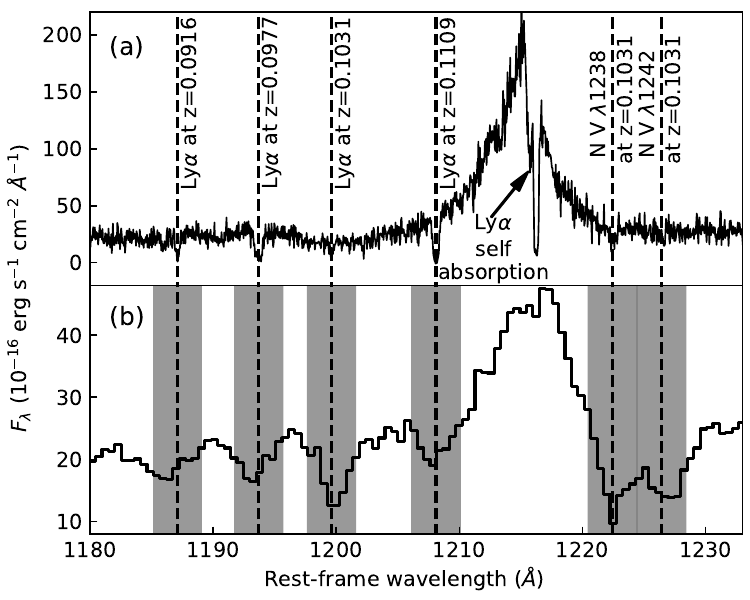}
  \caption{
  Identification of the UV absorption lines towards the nucleus of F01004-2237.
  {\bf (a)}: The \textit{HST}/COS spectrum and absorption lines identified in it.
  {\bf (b)}: We show the regions affected by the absorption lines in the STIS spectrum of the 2021 flare with grey shades.
  }
\label{fig:show_absline}
\end{figure}

We identify the absorption lines affecting the modelling of post-flare STIS spectrum with the help of an archival \textit{HST}/COS spectrum with a high spectral resolution.
The COS spectrum was taken on December 3, 2011 \citep[Proposal ID 12533, see][for more details]{Martin2015}.
We identified four absorption line systems in it (Fig.~\ref{fig:show_absline}(a)), whose redshifts are $0.110910\pm0.000013$, $0.103111\pm0.000046$, $0.097674\pm0.000027$ and $0.091610\pm0.000023$, respectively.
Absorption lines of Ly$\alpha$ and N V $\lambda$1238 were detected in the system at $z\sim0.1031$, while only Ly$\alpha$ was detected in the remaining three systems.
In addition, there are two absorption troughs in the Ly$\alpha$ emission line in the COS spectrum.
They may be originated in Ly$\alpha$ self-absorption of the inter-stellar medium as suggested by \cite{Martin2015}, and may have little effect on the emission of the 2021 flare from the galactic nucleus.

We calculated the wavelength ranges affected by the absorption lines in the STIS spectrum by assuming that their redshifts and widths did not change.
The absorption lines are intrinsically narrow, as can be seen from the COS spectrum.
Thus, their profiles in the STIS spectrum should be dominated by the line spread function (LSF).
According to the STIS Instrument Handbook\footnote{https://hst-docs.stsci.edu/stisihb/ }, when a G140L grating and a 0.2$\arcsec$ slit are used, the FWHM of the LSF near 1350 \AA\ is 1.6 pixels, corresponding to 210 km s$^{-1}$.
Considering the broad wing in the STIS LSF, we adopted $\pm500$ km s$^{-1}$ as the range each absorption line affects (Fig.~\ref{fig:show_absline}(b)).

\end{appendix}

\end{document}